\documentclass{article}

\pdfoutput=1

\usepackage{jheppub}
\usepackage{amsmath}
\usepackage{amssymb}
\usepackage{latexsym}
\usepackage{amsfonts}
\usepackage{graphicx}
\usepackage{epstopdf}

\usepackage{color}%
\usepackage{hyperref}
\def\hhref#1{\href{http://arxiv.org/abs/#1}{arXiv:#1}} 
\def\mc{\mathcal}
\def\bea{\begin{eqnarray}}
\def\ea{\end{eqnarray}}

\title{Resurgence and the Nekrasov-Shatashvili Limit: Connecting Weak and Strong Coupling in
the Mathieu and Lam\'e Systems}

\author[1]{G\"ok\c ce Ba\c sar}
\author[2]{and Gerald~V.~Dunne}
\affiliation[1]{Maryland Center for Fundamental Physics, University of Maryland, College Park, MD, 20742}
\affiliation[2]{Department of Physics, University of Connecticut, Storrs CT 06269}
\emailAdd{gbasar@umd.edu}
\emailAdd{gerald.dunne@uconn.edu}
\abstract{
The Nekrasov-Shatashvili limit for the low-energy behavior of  ${\mathcal N}=2$ and ${\mathcal N}=2^*$ supersymmetric $SU(2)$ gauge theories is encoded in the spectrum of the Mathieu and Lam\'e equations, respectively. This correspondence is usually expressed via an all-orders Bohr-Sommerfeld relation, but this neglects non-perturbative effects, the nature of which is very different in the electric, magnetic and dyonic regions.
In the gauge theory dyonic region the spectral expansions are divergent, and indeed are not Borel-summable, so they are more properly described by resurgent trans-series in which perturbative and non-perturbative effects are deeply entwined. 
In the gauge theory electric region the spectral expansions are convergent, but nevertheless there are non-perturbative effects due to poles in the expansion coefficients, and which we  associate with worldline instantons. This provides a concrete analog of a phenomenon found recently by Drukker, Mari\~no and Putrov in the large $N$ expansion of the ABJM matrix model, in which non-perturbative effects are related to complex space-time instantons.
In this paper we study how these very  different regimes arise from an exact WKB analysis, and join smoothly through the magnetic region. This approach also leads to a simple proof of a resurgence relation found recently by Dunne and \"Unsal, showing that for these spectral systems all non-perturbative effects are subtly encoded in perturbation theory, and identifies this with the Picard-Fuchs equation for the quantized elliptic curve.
}

\begin{document}                                                      
   
\date{\today}
\maketitle

\section{Introduction} 

In this paper we revisit the vacuum structure of $\mc N=2$ supersymmetric $SU(2)$ Yang-Mills theories using the resurgence formalism that unifies perturbative and non-perturbative physics\footnote{Here, perturbative and non-perturbative refers to the expansion in the Nekrasov deformation parameter of the gauge theory, not the gauge coupling.} 
\cite{Ecalle:1981,Costin:2009,Marino:2007te,Pasquetti:2009jg,Aniceto:2011nu,Marino:2012zq,Argyres:2012ka,Dunne:2012ae,Dunne:2013ada,Aniceto:2013fka,Aniceto:2014hoa,Dorigoni:2014hea}. The $\mc N=2$ supersymmetric $SU(2)$ Yang-Mills theory possesses a rich vacuum structure. The space of gauge inequivalent vacua, the moduli space, is a manifold parametrized by the scalar condensate $u  =\langle {\rm Tr}\, \Phi^2 \rangle$. In their seminal work, Seiberg and Witten showed that this manifold is precisely the moduli space of genus-one Riemann surfaces (i.e. tori) and the dynamics of the low energy effective theory can be formulated in a geometric language in the terms of elliptic curves \cite{Seiberg:1994rs,Gorsky:1995zq,Bilal:1995hc,Lerche:1996xu,AlvarezGaume:1996mv,Dorey:2002ik,Teschner:2014oja}. A further conceptual and computational breakthrough came with the introduction of the  Nekrasov partition function  \cite{Nekrasov:2002qd}, and the subsequent direct relation to integrable models and the Bethe ansatz \cite{Nekrasov:2009rc,Langmann:2014rja}.  Remarkably, the prepotential in the Nekrasov-Shatashvili limit is encoded in the spectra of certain states in certain simple Schr\"odinger systems through the monodromy and exact WKB properties of differential equations \cite{Mironov:2009uv,Fateev:2009aw,Maruyoshi:2010iu,He:2010xa,Huang:2011qx,Huang:2012kn,KashaniPoor:2012wb,Piatek:2013ifa,Krefl:2013bsa,Gorsky:2014lia}. The moduli parameter $u$ is directly identified with the eigenvalues of these Schr\"odinger systems, and the scalar (and dual scalar) field expectation values are identified with actions and dual actions in an all-orders WKB analysis.

In this paper we extend this approach to show that there are additional non-perturbative aspects of this relation that reflect the physics of the non-perturbatively small gaps and bands in the Schr\"odinger spectra. These spectra have three distinct physical regions, and these can be explicitly  associated with the three physical regimes, {\it electric}, {\it dyonic} and {\it magnetic},  of the supersymmetric (SUSY) gauge theory. The interplay of perturbation theory and non-perturbative physics is different in each region. The {\it dyonic} region is characterized by divergent perturbative expansions described by resurgent trans-series \cite{zjj,Dunne:2013ada,Basar:2013eka} that systematically unify perturbative and non-perturbative physics; the {\it electric} region has convergent perturbative expansions but there are nevertheless non-perturbative effects associated with poles of the expansion coefficients.  (This provides a concrete analog of a phenomenon found recently by Drukker, Mari\~no and Putrov \cite{Drukker:2010nc} in the large $N$ expansion of the ABJM matrix model, in which non-perturbative effects are related to complex space-time instantons, and which were subsequently related to poles in the 't Hooft expansion coefficients \cite{Hatsuda:2012hm}.) The {\it magnetic} region is a cross-over region in which non-perturbative effects are large, and in fact the spectral bands and gaps are of equal width. This correspondence between the Nekrasov-Shatashvili limit and monodromies of spectral problems also provides a simple proof of a surprising resurgence relation found by Dunne and \"Unsal in the spectra of certain quantum systems, which shows that all non-perturbative effects are subtly encoded in perturbation theory \cite{Dunne:2013ada}. 

An elementary but significant observation is that the energy eigenvalue, $u$, for the Schr\"odinger systems should be viewed as a function of {\bf two} variables, the coupling $\hbar$ and also the eigenvalue level label $N$: $u=u(N, \hbar)$. For a {\it uniform} analysis valid throughout the entire spectrum, it is natural to define a ``'t Hooft parameter''\footnote{Note that our $N$ is not $N_c$. In this paper we are discussing $SU(2)$ gauge theory. The role of $N_c$ (or the $N$ of the matrix model) is played in this context by the level number $N$. It is also possible to introduce yet another parameter, $N_c$, from the $SU(N_c)$ Toda system, as in \cite{Mironov:2009uv}, but this is not done in the current paper.}, 
\begin{eqnarray}
\lambda\equiv N \hbar
\label{thooft}
\end{eqnarray}
and consider different limits of the two parameters, including a double-scaling limit. The usual analyses of divergent perturbative expansions \cite{LeGuillou:1990nq,zjj,Dunne:2013ada}, 
in the $\hbar \to 0$ limit, and their associated resurgent trans-series representations, are implicitly restricted to a particular non-uniform limit in which the eigenvalue level number $N$ is small, $N\ll \frac{1}{\hbar}$, corresponding to states with energy well below the energy barrier.  More generally, the perturbative expression for the $N^{\rm th}$ energy eigenvalue has the form
\begin{eqnarray}
u(N, \hbar) = u(N, \lambda) =\sum_{n=0}^\infty N^{2-2n} F_n(\lambda) =\sum_{n=0}^\infty \hbar^{2n-2} \tilde{F}_n(\lambda)
\label{eq:genus}
\end{eqnarray}
which is analogous to the genus expansion of the free energy for a matrix model or gauge theory system \cite{Marino:2012zq}. We show that there are in fact non-perturbative corrections to this expression, of the form $e^{-N/\lambda}$ for small $\lambda$, but of the form $e^{-2 N \,\ln \lambda}$ for large $\lambda$ (see (\ref{eq:smallq-splitting})). Moreover, in this large $\lambda$ regime the perturbative expansions are convergent, but the perturbative coefficients have poles that are responsible for the non-perturbative splittings. This can be compared with recent results  concerning non-perturbative contributions to matrix models associated with the ABJM free energy \cite{Drukker:2010nc,Hatsuda:2012hm,Kallen:2013qla}, where in the large $N$ limit there are extra non-perturbative terms of the form $e^{-N/\sqrt{\lambda}}$ [$\lambda$ is the  't Hooft coupling], and these extra terms are related to poles of the expansion coefficients \cite{Hatsuda:2012hm,Kallen:2013qla}.

We first review some basic facts from the SUSY gauge theory side \cite{Seiberg:1994rs,Gorsky:1995zq,Bilal:1995hc,Lerche:1996xu,AlvarezGaume:1996mv,Dorey:2002ik} of the correspondence and set our notation, in order to make the precise identification between the gauge theory quantities and the corresponding spectral quantities. The vacuum expectation values of the scalar field and its dual partner, $a_0(u)$ and $a_0^D(u)$, correspond to the two independent cycles on the torus given as
\bea
a_0(u)&=& \oint_{\gamma_1} \mu \equiv \frac{\sqrt{2}}{\pi} \int_0^\pi \sqrt{u-\Lambda^2\cos\phi}\,d\phi 
\label{SW-periods1}\\
a_0^D (u) &=& \oint_{\gamma_2} \mu \equiv \frac{\sqrt{2}}{\pi} \int_0^{\cos^{-1}(u/\Lambda^2)} \sqrt{u-\Lambda^2\cos\phi}\,d\phi 
\label{SW-periods2}
\ea
where $\Lambda$ is the dynamically generated scale, and $\gamma_1, \gamma_2$ are integration cycles discussed in detail below, in Section III.  In this form $a_0(u)$ and $a_0^D$ correspond to lowest order WKB cycles for the Mathieu system, hence the subscripts "0" \cite{Gorsky:1995zq}. The information about the BPS states of the theory is also contained in these cycles, where the central charge and mass of a BPS state are given as
\bea
Z_{q_e,q_m}(u)=q_e a_0(u)+q_m a^D_0(u)\,,\qquad M_{q_e,q_m} =\sqrt{2}\,|q_e a_0(u)+q_m a^D_0(u)|\,.
\ea
Here $q_e$ and $q_m$ are integers that denote the electric and magnetic charge of the state.

There are three singular points in the moduli space where one or both of the cycles acquire branch points. They are $u\rightarrow\infty$, $u=\Lambda^2$ and $u=-\Lambda^2$. Around these points on the moduli space the low energy theory is described by weakly coupled massive $Z$ bosons, almost massless magnetic monopoles and almost massless dyons, respectively. For the rest of the paper we will refer to the local neighborhoods around these points as the electric, magnetic and dyonic regions, respectively. In general, there are two separate sectors in the moduli space with different particle spectra. In one sector, the spectrum of the theory consists of $Z$ bosons and an infinite tower of dyonic tower with charges $\pm(q_e,1)$. In the other, it consists of magnetic monopoles with charge $\pm(0,1)$, and dyons with $\pm(1, \pm1)$. These two regions are separated by a closed curve, $\mc Im[a_0^D/a_0]=0$, called the {\it curve of marginal stability}, where the BPS states with higher charges can decay into monopoles and dyons. This curve is approximately an ellipse around $u=0$ and contains the points $u=\pm\Lambda^2$. 

 Another important object for the low energy effective theory is the  {\it prepotential}, $\mc F_0(a_0)$, which is a holomorphic function whose derivatives are 
 \bea
\frac{\partial \mc F_0}{\partial a_0}=a_0^D,\qquad \frac{\partial^2 \mc F_0}{\partial a_0^2}=\frac{\theta_{YM}(a_0)}{2\pi}+i \frac{4 \pi}{g^2_{YM}(a_0)}\,.
 \ea
where $\theta_{YM}$ and $g_{YM}$ are the theta parameter and the coupling constant (at the scale $a_0$) of the gauge theory. In the electric region where $a_0\gg \Lambda$, the prepotential has the following semiclassical expansion:
\bea
\mc F_0(a_0)&=& \mc F_0^{class.}(a_0)+\mc F_0^{pert.}(a_0)+\mc F_0^{inst.}(a_0)\nonumber\\
&=&\frac{1}{2}\tau_0 a_0^2+ i\frac{a_0^2}{2\pi} \log\left(\frac{a_0^2}{\Lambda^2}\right)+\frac{a_0^2}{2 \pi i}\sum_{k=1}^\infty c_{0,k}\left(\frac{\Lambda}{a_0}\right)^{4k}
\label{SW-prepotential}
\ea
The first two terms in this expansion are the classical and one-loop contributions, and the last term is the sum over non-perturbative $k$-instanton corrections\footnote{A comment here concerning terminology: Throughout the paper, we use the word ``instanton'' to describe two different objects: (i) the BPST instantons that appear in the 4d gauge theory; (ii) the quantum mechanical instantons that appear in the quantum mechanical description of the deformed gauge theory. The distinction should be apparent within the context of the discussion, and for the bulk of the paper it will mostly refer to (ii).}
The first few terms of the instanton expansion calculated via standard field theory methods agree with the extraction of the coefficients from the Seiberg- Witten solution.
For example, for the $\mathcal N=2$ SUSY $SU(2)$ gauge theory, the instanton expansion in \eqref{SW-prepotential} is 
\bea
\mc F_0^{inst.}(a_0)&=&{\Lambda^4\over 8 \pi i a_0^2} + {5\Lambda^8\over256 \pi i a_0^6}  + {3\Lambda^{12}\over256 \pi i a_0^{10}}  + {1469\Lambda^{16}\over2^{31} \pi i a_0^{14}} +\dots
\nonumber\\
&=&= \frac{a_0^2}{2 \pi i}\sum_{k=1}^\infty c_{0,k}\left(\frac{\Lambda}{a_0}\right)^{4k}\,.
\label{SW-prepotential_inst}
\ea
However beyond the two-instanton level,  the direct quantum field theory methods become computationally very difficult.

An alternative way to calculate the instanton sum is through localization \cite{Nekrasov:2002qd,Nekrasov:2009rc}. In this approach, a two-parameter generalization of the prepotential, $\mc F(a|\epsilon_1,\epsilon_2)$, is introduced. The parameters characterize certain SUSY preserving space-time deformations. The prepotential of the deformed theory is calculable via localization technique and one obtains the Seiberg-Witten prepotential $\mc F_0(a)$ in the limit
\bea
\mc F_0(a)=\lim_{\epsilon_1,\epsilon_2\rightarrow0} \mc F(a|\epsilon_1,\epsilon_2)\,.
\ea 
In this paper we will focus on the Nekrasov-Shatashvili limit \cite{Nekrasov:2009rc}, a particular one parameter deformation of the gauge theory with $\epsilon_1\equiv\hbar$, and $\epsilon_2=0$. We denote the prepotential in this limit as 
\bea
\mc F_{NS}(a,\hbar)&=&\lim_{\epsilon_2\rightarrow0} \mc F(a|\epsilon_1=\hbar,\epsilon_2)\nonumber\\
&=&\mc F^{class.}(a,\hbar)+\mc F^{pert.}(a,\hbar)+\mc F^{inst.}(a,\hbar)\nonumber\\
&=&\sum_{n=0}^\infty \hbar^{2n} \mc F_n(a)\,.
\ea
Continuing the example (\ref{SW-prepotential_inst}) of $\mathcal N=2$ SUSY $SU(2)$ gauge theory, 
the expansions for the perturbative and instanton contributions are
\bea
\mc F^{inst.}(a,\hbar)&=& \mc F_0^{inst.}(a)+{\hbar ^2\over2 \pi i} \left(\frac{\Lambda ^4}{16 a ^4}+\frac{21 \Lambda ^8}{256 a ^8}+\dots\right)+{\hbar ^4\over 2 \pi i} \left(\frac{\Lambda ^4}{64 a ^6}+\frac{219 \Lambda ^8}{2048 a ^{10}}+\frac{1495 \Lambda ^{12}}{3072 a ^{14}}\dots\right)+\dots \nonumber \\
\mc F^{class.}(a,\hbar)&+&\mc F^{pert.}(a,\hbar)=
\text{(quadratic poly. in $a$)}-{a^2\over 2\pi i}\log{a^2\over\Lambda^2}-\frac{\hbar^2}{48 \pi i}\log\frac{a ^2}{2\Lambda ^2}+\hbar^2 \sum_{n=1}^\infty d_{2n}\left({\hbar\over a}\right)^{2n}\nonumber\\
 \label{eq:ns-prepotential}
\ea

In the Nekrasov-Shatashvili limit, there is a direct correspondence between the gauge theory and integrable models \cite{Nekrasov:2009rc,Langmann:2014rja}, and also with 0+1 dimensional (i.e. quantum mechanical) Sine-Gordon theory for $SU(2)$ $\mc N=2$ SUSY gauge theory, and Lam\'e theory for $SU(2)$ $\mc N=2^*$ SUSY gauge theory, a model with a massive hypermultiplet in the adjoint representation \cite{Mironov:2009uv,Fateev:2009aw,Maruyoshi:2010iu,He:2010xa,Huang:2011qx,Huang:2012kn,KashaniPoor:2012wb,Piatek:2013ifa,Krefl:2013bsa}. In the limit where the mass of the hypermultiplet, $m$, is zero, the $\mc N=2^*$ theory becomes $\mc N=4$ theory, while at large mass it reduces back to the $\mc N=2$ theory.  In particular, the gauge theory moduli parameter, $u$, is encoded in the energy eigenvalue of the quantum mechanical (QM) system described by the time independent Schr\"odinger equation
\bea
\mc N=2 & : & \qquad -\frac{\hbar^2}{2}\,\frac{d^2\psi}{dx^2}+ \cos(x)\, \psi=u\, \psi \\
\mc N=2^* & : & \qquad -\frac{\hbar^2}{2}\,\frac{d^2\psi}{dx^2}+\frac{1}{8}\left(m^2-\frac{\hbar^2}{4}\right)  {\mc P}\left(\frac{x}{2}+{i\pi {\mathbb K}^\prime\over{\mathbb K}}; \tau\right)\psi=u\, \psi
\label{eq:schrodinger}
\ea
where $\mc P$ is the Weierstrass elliptic function. Some minor rescaling is required [as discussed in Section \ref{sec:lame-electric}] to have a smooth decoupling limit in which the $\mathcal N=2^*$ theory reduces to the $\mathcal N=2$ theory, taking $m^2\to\infty$ limit, combined with vanishing elliptic parameter $k^2\to 0$, such that $m^2 k^2$ is finite.

Furthermore, at nonzero $\hbar$, the periods (\ref{SW-periods1}, \ref{SW-periods2}) generalize to the Bohr-Sommerfeld integrals of the QM system \cite{Mironov:2009uv}. They can be expressed formally as all orders WKB expansions:
\begin{eqnarray}
a(u, \hbar)=\sum_{n=0}^\infty \hbar^{2n} \, a_{n}(u)\qquad, \qquad a^D(u, \hbar)=\sum_{n=0}^\infty \hbar^{2n} \, a_{n}^D(u)\,.
\label{eq:periods}
\end{eqnarray}
One recovers the Seiberg-Witten solution in the continuum limit obtained from $\hbar=0$,  the leading order WKB approximation to \eqref{eq:periods}. However, this is only part of the story, as these Bohr-Sommerfeld expressions (\ref{eq:periods}) give only the {\it locations} of the bands or gaps. There is also important spectral information in the non-perturbative {\it widths} of bands and gaps. This information is connected to the perturbative information in quite different ways in different parts of the spectrum, corresponding to different semiclassical behavior for the different regions of the 't Hooft parameter $\lambda$. Using the direct correspondence, we can associate this with different physical behavior in the corresponding SUSY gauge theory.

The relation between the Nekrasov-Shatashvili limit and exact quantization conditions of quantum mechanical systems has also been discussed recently in   \cite{Krefl:2013bsa}, from the perspective of holomorphic $\beta$-ensembles and the associated quantum geometry \cite{Aganagic:2011mi}. Here we follow a complementary approach, building our analysis on elementary WKB methods, treated {\it uniformly} across the entire spectrum. This relation has also been formulated in terms of Whitham dynamics \cite{Gorsky:2014lia} applied to exact quantization conditions for complex quantum mechanical systems.

In this paper we build a unified WKB analysis that spans all regions of the spectrum, using  all-orders exact WKB analysis  \cite{dunham,bender,voros,takei,delabaere,howls-book}.
In Section II we review basic properties of the Mathieu spectrum, and its relation to the Nekrasov partition function in the Nekrasov-Shatashvili limit. Section III contains the uniform all-orders WKB analysis. In Section IV we describe a physical analog of the transition between different spectral regions, in terms of the transition between tunneling and multi-photon pair production in the Schwinger effect \cite{Heisenberg:1935qt,Schwinger:1951nm,Dunne:2004nc,keldysh,brezin,popov}. 
In Section V we discuss a Picard-Fuchs interpretation that explains the significance of resurgence, and also show that  this SUSY gauge theory perspective in terms of the Nekrasov-Shatashvili limit yields a simple proof of the recently found Dunne-\"Unsal relation that connects the fluctuations about the perturbative vacuum with the fluctuations about the one-instanton sector, and with all higher multi-instanton sectors \cite{Dunne:2013ada}. 
The Lam\'e system is discussed in Section VI, by means of a complementary approach to the large $\lambda$ region, based on the Gelfand-Dikii expansion of the resolvent. We end with a summary and comments about future work.

\section{Mathieu Equation and $SU(2)$ $\mathcal N=2$  SUSY Gauge Theory}

\subsection{Mathieu Equation: Notation and Basic Spectral Properties}

In this Section we review relevant facts about the spectrum of the Mathieu equation \cite{nist,meixner,muller,magnus}, translated into notation that makes explicit the relation to the Nekrasov partition function. 
The standard textbook form of the Mathieu equation is (\url{http://dlmf.nist.gov/28}):
\begin{eqnarray}
\psi^{\prime\prime}+\left(A-2Q\, \cos(2z)\right)\psi=0\quad\longrightarrow\quad -\frac{\hbar^2}{2}\,\frac{d^2\psi}{dx^2}+\Lambda^2 \cos(x)\psi=u\, \psi
\label{mathieu1}
\end{eqnarray}
(We use capital letters $A$ and $Q$, rather than  the conventional lower-case ones, as the symbols $a$ and $q$ have special meaning in the gauge theory discussion).
We thus make  the  identifications:
\begin{eqnarray}
Q=\frac{4 \Lambda^2}{\hbar^2}\qquad, \qquad A=\frac{8 \,u}{\hbar^2}
\label{hbar}
\end{eqnarray}
We will mostly set the scale $\Lambda=1$ in what follows, re-introducing it where necessary by simple dimensional scaling arguments.

\begin{figure}[htb]
\centering
\includegraphics[scale=.5]{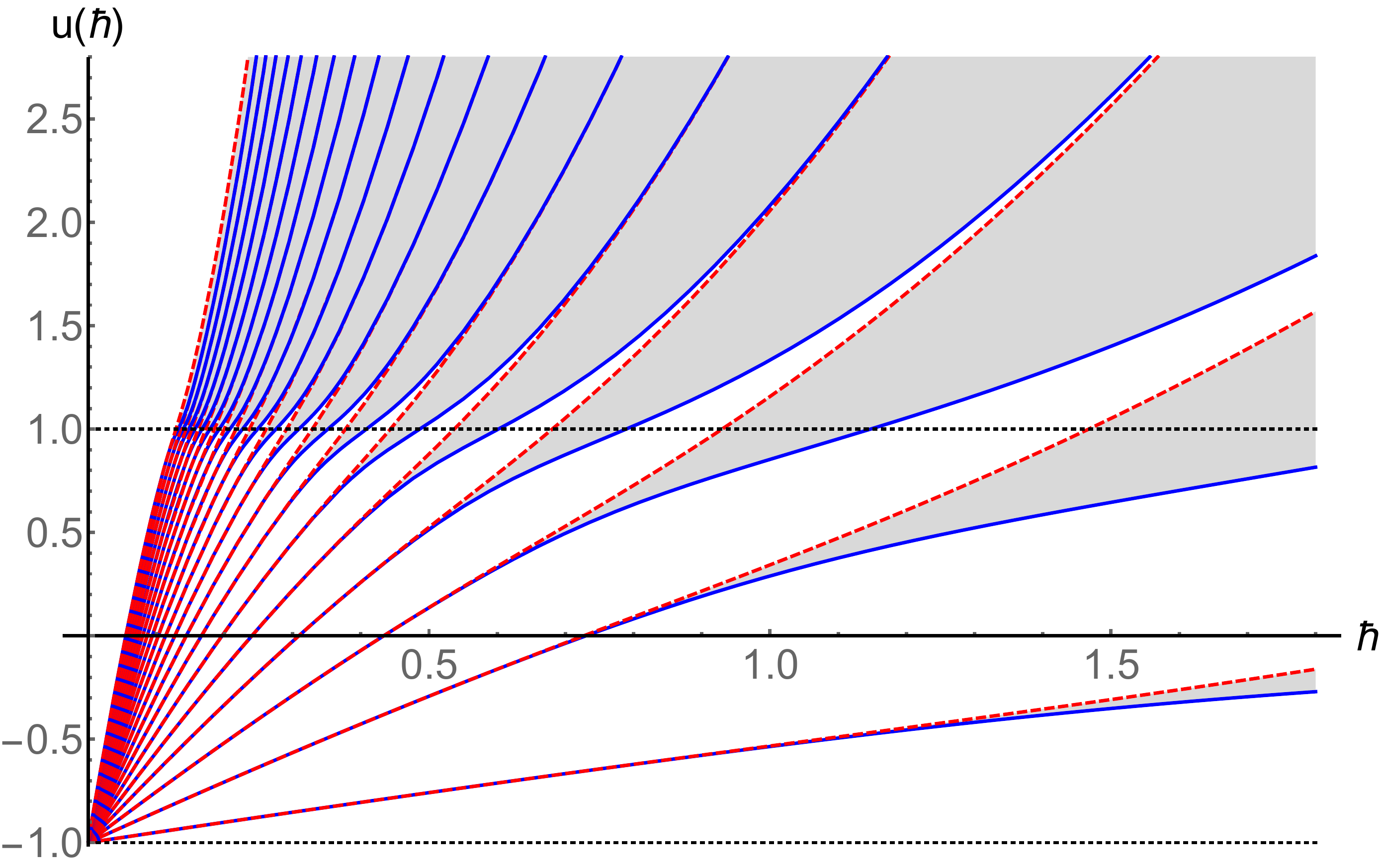}
\caption{The band spectrum of the Mathieu equation, expressing the eigenvalue $u$ as a function of the parameter $\hbar$, as  in (\ref{mathieu1}). This eigenvalue $u$ will be identified with the  scalar condensate moduli parameter in the SUSY gauge theory.  The bands are shaded in grey, with the lower edges of each band shown as a solid (blue) line, and the top edge of each band as a dashed (red) line. At small $\hbar$, the bands are exponentially narrow, and the band {\it location} follows the linear behavior in (\ref{eq:largeq}). At large $\hbar$ the gaps are exponentially narrow, and the gap {\it location} follows the quadratic behavior in (\ref{eq:smallq}). 
The top and bottom of the potential, at $u=\pm 1$, are shown as dotted lines.
Notice the smooth transition between exponentially narrow bands (shaded) at  small $\hbar$, and exponentially narrow gaps (unshaded) at large $\hbar$. This transition occurs at the top of the potential, where $u=1$, shown as a straight line. Note that in the vicinity of the barrier top, the bands and gaps are of equal width, and are not exponentially narrow, as discussed Section \ref{sec:wkb-magnetic}.
}
\label{fig:f1}
\end{figure}
\begin{figure}[htb]
\centering
\includegraphics[scale=.5]{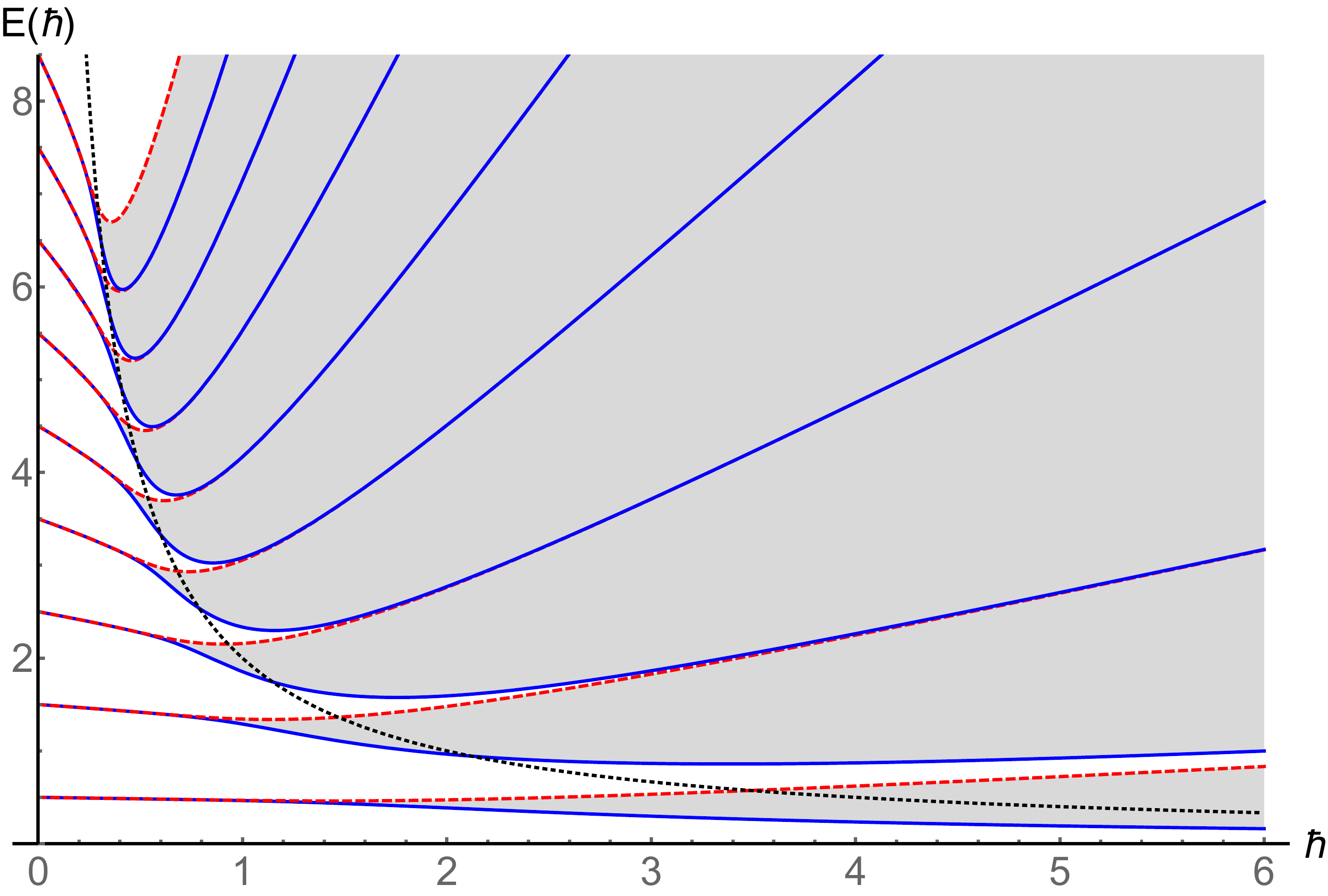}
\caption{Another view of the band spectrum of the Mathieu equation, expressed here in terms of the rescaled eigenvalue $E$, as in  (\ref{hbar-g}), and with the scale $\Lambda=1$. The rescaled eigenvalue $E$ is to be compared with the normalizations in \cite{zjj,Dunne:2013ada}. The transition region between exponentially narrow bands (small $\hbar$) and exponentially narrow gaps (large $\hbar$) occurs at the top of the potential, where $E=2/\hbar$, as shown as a dotted  line. Note that in the vicinity of the barrier top, the bands and gaps are of equal width, and are not exponentially small.}
\label{fig:f2}
\end{figure}
We also make explicit comparison with the work of Zinn-Justin and Jentschura \cite{zjj} (see also \cite{Dunne:2013ada}), who used a different scaling:
\begin{eqnarray}
\left(-\frac{1}{2}\frac{d^2}{dx^2}+\frac{1}{8 g}\sin^2(2 \sqrt{g}\, x)\right)\psi= E_{ZJJ}\, \psi
\quad\longrightarrow\quad 
\left(-\frac{(16g)^2}{2}\frac{d^2}{dx^2}+\cos(x)\right)\psi=(16g\, E_{ZJJ}-1)\, \psi \nonumber \\
\label{zj}
\end{eqnarray}
Thus,  we identify (note: we flipped the sign of $\cos(x)$ by a simple half-period shift)
\begin{eqnarray}
\hbar =16\, g \qquad, \qquad u=-1+16\,g\, E_{ZJJ} =-1+\hbar \, E_{ZJJ}
\label{hbar-g}
\end{eqnarray}
Because of the direct identification of the rescaled eigenvalue $u$ with the SUSY gauge theory scalar condensate,  we will describe the Mathieu spectrum in terms of this eigenvalue $u$, with conversions to $A$ or $E$  made using the above re-scalings (\ref{hbar}, \ref{zj},  \ref{hbar-g}).

The exact Bloch spectral condition, or Floquet analysis \cite{nist,meixner,magnus},  can be expressed in terms of two independent solutions, $\psi_1(z;  u, \hbar)$ and $\psi_{2}(z; u, \hbar)$, normalized at $z=0$ as:
\begin{eqnarray}
\begin{bmatrix}\psi_{1}(0; u, \hbar)&\psi_{2}(0; u, \hbar)\\
\psi^{{\prime}}_{1}(0; u, \hbar)&\psi^{{\prime}}_{2}(0; u, \hbar)%
\end{bmatrix}=\begin{bmatrix}1&0\\
0&1\end{bmatrix}.
\end{eqnarray}
The Bloch boundary condition, $\psi(z+\pi)=e^{{i\theta}}\,\psi(z)$, can then be written in compact form in terms of $\psi_1(\pi; u, \hbar)$ evaluated one half period away from the normalization point:
\begin{eqnarray}
\cos\left(\theta\right)=\psi_{1}(\pi;u, \hbar)
\label{bloch}
\end{eqnarray}
Equation (\ref{bloch})  is the  ``exact quantization condition'', implicitly expressing the eigenvalue $u$ in terms of $\hbar$ [and therefore $E$, or $A$, in terms of $g$, or $Q$, using the re-scalings (\ref{hbar}, \ref{zj},  \ref{hbar-g})], for each Bloch parameter $\theta$. The band/gap edges correspond to $\theta$ being an integer multiple of $\pi$. However the exact quantization condition (\ref{bloch}) is of limited practical use unless we have an explicit expression for the normalized Mathieu function $\psi_1$.  Concrete approximations for the eigenvalues can be obtained from  the exact quantization condition  by making expansions of the Mathieu functions in terms of other functions, such as trigonometric or Hermite functions \cite{meixner,magnus,muller}. Different expansions are suitable for different regions in the spectrum, as is familiar from elementary solid state physics. Deep inside the wells, we use the tight-binding approximation in terms of `atomic' states bound in the wells, while far above the barrier we use the `neary-free-electron model'  \cite{peierls}.

The spectrum of the Mathieu equation consists of an infinite sequence of bands and gaps, as shown in Figures \ref{fig:f1} and \ref{fig:f2}. 
In Figures \ref{fig:f1} and \ref{fig:f2}, we see the transition from exponentially narrow bands low in the spectrum, to exponentially narrow gaps higher in the spectrum. The cross-over occurs near the top of the potential, where $u\sim 1$. In fact, this transition occurs when $N\hbar \sim\frac{8}{\pi}$, as discussed in Section \ref{sec:wkb-magnetic}. In this transition region the bands and gaps are of equal width, and neither is exponentially narrow.  In Figure \ref{fig:f1} we see a transition from narrow bands, with locations approximately linear in $N$ and $\hbar$, to narrow gaps, with locations approximately quadratic in $N$ and $\hbar$. 
Figure \ref{fig:f2} illustrates the same behavior, in terms of the rescaled energy eigenvalue $E=(u+1)/\hbar$, which emphasizes the transition from harmonic oscillator behavior at small $\hbar$, to particle-on-a-circle behavior at large $\hbar$.
\begin{figure}[htb]
\centering
\includegraphics[scale=.4]{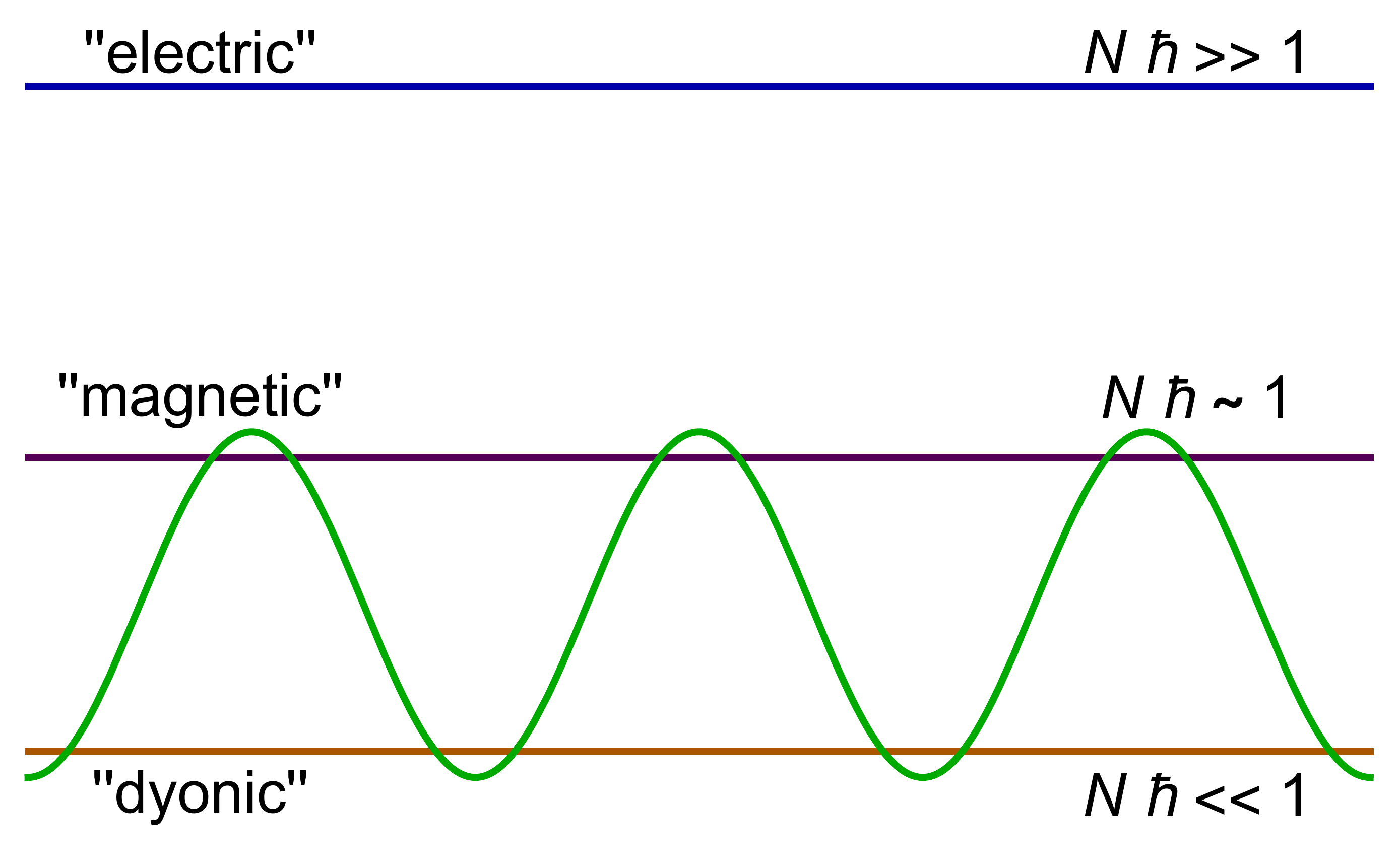}
\caption{Three different spectral regimes for the Mathieu equation: 
(i) far above the barrier, where $u\gg 1$, and $N \hbar\gg 1$; 
(ii) barrier top region, where $u\sim 1$, and $N \hbar\sim 1$; 
(iii) deep inside the wells, where $u\sim -1$, and $N \hbar \ll 1$. 
In gauge theory language, these are the ``electric'', ``magnetic'' and ``dyonic'' regions, respectively, as shown in the Figure.}
\label{fig3}
\end{figure}

There are clearly three interesting spectral regimes, which we identify with three interesting physical regions of the gauge theory, as shown in Figure \ref{fig3}:
\begin{itemize}

\item dyonic: deep inside potential wells: $u\sim  -1$, $N\hbar \ll 1$, exponentially narrow bands

\item magnetic: near the barrier top: $u\sim +1$, $N\hbar \sim 1$, bands and gaps of equal width

\item  electric: high above barrier top:  $u\gg 1$, $N\hbar \gg 1$, exponentially narrow gaps

\end{itemize}

In terms of the `` 't Hooft coupling'' (\ref{thooft}), $\lambda\equiv  N\, \hbar$, we can consider the energy eigenvalue as a function of $N$ and $\hbar$, or $N$ and $\lambda$:
\begin{eqnarray}
u(N, \hbar)=u(N, \lambda)
\label{eq:2var}
\end{eqnarray}
The semiclassical limit, $\hbar\to 0$, with $\lambda$ fixed, requires requires $N\to \infty$; but we still have three different regions, where $\lambda\ll 1$, $\lambda\sim 1$, or $\lambda\gg 1$. (In fact, we show later that the cross-over region is at $\lambda \sim \frac{8}{\pi}$). A {\it uniform} analysis, valid for all $\lambda$, therefore permits access to all regions of the spectrum. For example, the small $\lambda$ region can be interpreted as $\hbar\to 0$, with $N$ fixed, which therefore gives information about the weak-coupling expansion for low-lying energy levels with $N \ll \frac{1}{\hbar}$. On the other hand, the large $\lambda$ region gives information about the strong-coupling expansion for low-lying modes, with $N$ fixed and $\hbar\gg \frac{1}{N}$.

There is an interesting inversion of the meaning of strong and weak coupling. In gauge theory language, the electric region is weakly coupled, while the dyonic  region is strongly coupled. On the other hand, in quantum mechanical language in the scaling (\ref{zj}) of \cite{zjj}, in terms of the coupling $g$, the situation is reversed.

\subsection{Dyonic Region: Resurgent Trans-series Expansions, Deep Inside the Wells}
\label{sec:dyonic}

In this $\lambda\ll 1$ regime, which is weak-coupling in the QM sense but strongly-coupled in the gauge theory sense, the perturbative expansions for the energy levels are divergent. The formal perturbative expansion reads (\url{http://dlmf.nist.gov/28.8.E1}), translated into our notation:
 \begin{eqnarray}
 u(N, \hbar) 
&\sim& -1 +\hbar \left[N+\frac{1}{2}\right] -\frac{\hbar^2}{16}\left[\left(N+\frac{1}{2}\right)^2+\frac{1}{4}\right]-\frac{\hbar^3}{16^2}\left[\left(N+\frac{1}{2}\right)^3+\frac{3}{4}\left(N+\frac{1}{2}\right)\right] \nonumber\\
&&  - \frac{\hbar^4}{16^3}\left[ \frac{5}{2}\left(N+\frac{1}{2}\right)^4+\frac{17}{4}\left(N+\frac{1}{2}\right)^2+\frac{9}{32}\right]  \nonumber\\
&& 
-  \frac{\hbar^5}{16^4}\left[ \frac{33}{4}\left(N+\frac{1}{2}\right)^5+\frac{205}{8}\left(N+\frac{1}{2}\right)^3+\frac{405}{64}\left(N+\frac{1}{2}\right)\right] -\dots
\label{eq:largeq}
\end{eqnarray}
For fixed level number $N$, with $N\ll 1 /\hbar$, the expansions for $E\equiv \frac{u+1}{\hbar}$ are factorially divergent and non-alternating, as series in 
$\hbar$ \cite{zjj,Dunne:2013ada}:
\begin{eqnarray}
E(N, g)=\sum_{n=0}^\infty c_n(N) \hbar^n\qquad, \quad c_n(N)\sim -\frac{2^{2N}}{\pi 16^{n+2N+1}}\frac{\Gamma(n+2N+1)}{(N!)^2}
\label{eq:large-order}
\end{eqnarray} 
These perturbative expansions are therefore non-Borel-summable, and so are incomplete on their own, and should be extended to real, unambiguous trans-series expansions \cite{zjj,Dunne:2013ada}:
\begin{eqnarray}
u(N, \hbar) =
\sum_{n=0}^\infty\sum_{k=0}^\infty \sum_{l=0}^{k-1} c_{n k l}(N)\, \hbar^n \left[\frac{\exp\left(- \frac{8}{\hbar}\right)}{\hbar^{N-1/2}}\right]^k \left[ \ln \left(\frac{- 1}{\hbar}\right)\right]^l 
\label{eq:trans}
\end{eqnarray}
The trans-series is an expansion in terms of instanton factors, each multiplied by fluctuations, and also by powers of logarithms due to quasi-zero modes coming from instanton/anti-instanton interactions \cite{Bogomolny:1980ur,ZinnJustin:1981dx,Balitsky:1985in}. 
Note that these log terms first appear at the 2-instanton level.
The trans-series coefficients $c_{n k l}(N)$ are related to one another in intricate ways, such that all imaginary parts cancel, leaving a real and unambiguous energy \cite{zjj,Dunne:2013ada}. Mathematically speaking, the advantage of the trans-series is that it encodes {\bf all} information about the function being computed, and is rigorously equivalent to the function wherever the function exists \cite{Ecalle:1981,Costin:2009}, in contrast to an asymptotic perturbative expansion such as (\ref{eq:largeq}).

The resurgent trans-series  incorporates non-perturbative contributions at all multi-instanton orders, the lowest of which is the exponentially small band width:
\begin{eqnarray}
\Delta u_{N}^{\rm band} 
 \sim  \sqrt{\frac{2}{\pi}} \frac{2^{4(N+1)}}{N!} \left(\frac{2}{\hbar}\right)^{N-1/2}
\exp\left[-\frac{8}{\hbar}\right]\left\{ 1-\frac{\hbar}{32}\left[3 \left(N+\frac{1}{2}\right)^2+4 \left(N+\frac{1}{2}\right) +\frac{3}{4}\right]+O(\hbar^2) \right\} \nonumber \\
\label{eq:bandwidth}
\end{eqnarray}
This band splitting is a {\it single-instanton} effect, and is real and unambiguous. The factor $8$ in the exponent is the instanton action: $\sqrt{2}\int_{-\pi}^\pi\sqrt{\cos(x)+1}\, dx =8$.

However, the non-Borel-summable perturbative series in (\ref{eq:largeq}) diverges at a rate associated with the two-instanton sector [more precisely, the instanton/anti-instanton part thereof], and lateral Borel summation produces an ambiguous imaginary non-perturbative term $\sim \pm i\, \exp\left[-16/\hbar\right]$. This ambiguous imaginary non-perturbative term is in fact cancelled by an identical term coming from an instanton gas analysis of the instanton/anti-instanton interaction \cite{zjj,Dunne:2013ada}. This leading cancellation, at the two-instanton level, is just the tip of the iceberg:  the cancellations between imaginary terms produced by lateral Borel summation of perturbation theory, and those coming from the multi-instanton sectors, occur at all orders, and these cancellations are encoded in relations between the coefficients of the resurgent trans-series expansion.

For example \cite{Dunne:2013ada}, for the lowest band, the large-order behavior of the perturbative coefficients is:
\begin{eqnarray}
c_n(0)\sim n!\left({\color{blue}1}-{\color{blue}\frac{5}{2}}\cdot\frac{1}{n}-{\color{blue}\frac{13}{8}}\cdot\frac{1}{n(n-1)}-\dots\right)
\label{eq:sg-large-order}
\end{eqnarray}
while the fluctuations about the instanton/anti-instanton saddle are given by
\begin{eqnarray}
{\rm Im}\, E(0, \hbar) &\sim& \pi\, e^{-16/\hbar}\left({\color{blue}1}-{\color{blue}\frac{5}{2}} \cdot \left(\frac{\hbar}{16}\right)^2-
{\color{blue}\frac{13}{8}} \cdot \left(\frac{\hbar}{16}\right)^4-\dots\right)
\label{eq:sg-iibar}
\end{eqnarray}
Note the precise correspondence of the coefficients in these two very different expansions. Equation (\ref{eq:sg-large-order})  is associated with the fluctuations about the perturbative vacuum saddle point, while Equation (\ref{eq:sg-iibar}) describes the fluctuations about the nonperturbative instanton/anti-instanton saddle point, the smallest action saddle-point with the same vacuum quantum numbers. This is an explicit example of the relations between coefficients that exist within the trans-series, and a direct manifestation of  resurgence.  In path integral language, resurgence  means that the fluctuations about various different saddle points in the multi-instanton expansion are directly related to one another. Such resurgent relations persist to all orders of the non-perturbative and quasi-zero-mode expansions \cite{Dunne:2013ada}.

Zinn-Justin and Jentschura (ZJJ) \cite{zjj} have argued the remarkable result that the entire trans-series can be generated from an exact quantization condition, together with just two functions: $B_{ZJJ}(E, g)$, which describes the perturbative series, and $A_{ZJJ}(E, g)$, which effectively describes the fluctuations around the single-instanton. In fact, the resurgent trans-series structure follows naturally from a uniform WKB analysis, which shows it is an expression of the analytic continuation properties of the parabolic cylinder functions \cite{Dunne:2013ada}. The ZJJ exact quantization condition is written as \cite{zjj} (we rewrite the following expressions in terms of $\hbar \equiv 16 g$ instead of $g$):
\begin{eqnarray}
\left(\frac{32}{\hbar}\right)^{-B_{\rm ZJJ}}\frac{e^{\frac{1}{2}A_{\rm ZJJ}}}{\Gamma\left(\frac{1}{2}-B_{\rm ZJJ}\right)}+
\left(-\frac{32}{\hbar}\right)^{-B_{\rm ZJJ}}\frac{e^{-\frac{1}{2}A_{\rm ZJJ}}}{\Gamma\left(\frac{1}{2}+B_{\rm ZJJ}\right)}=\frac{2\, \cos\theta}{\sqrt{2\,\pi}}
\label{eq:zjj-exact}
\end{eqnarray}
where $\theta$ is the Bloch angle, and for  the Mathieu potential, the two functions $B_{\rm ZJJ}(E, \hbar)$ and $A_{\rm ZJJ}(E, \hbar)$ are given by:
\begin{eqnarray}
B_{\rm ZJJ}(E, \hbar)&=&E+\frac{\hbar}{16}\left(\frac{1}{4}+E^2\right) 
+\left(\frac{\hbar}{16}\right)^2 \left(\frac{5E}{4}+3 E^3\right) 
+\left(\frac{\hbar}{16}\right)^3\left(\frac{17}{32}+\frac{35 E^2}{4}+\frac{25 E^4}{2}\right) \nonumber\\
&& +\left(\frac{\hbar}{16}\right)^4\left(\frac{721 E}{64}+\frac{525 E^3}{8}+\frac{245 E^5}{4}\right) +\dots
\label{eq:zjjb}\\
A_{\rm ZJJ}(E, \hbar)&=&\frac{16}{\hbar}+\frac{\hbar}{16} \left(\frac{3}{4}+3 E^2\right)
+\left(\frac{\hbar}{16}\right)^2 \left(\frac{23 E}{4}+11 E^3\right) 
 +\left(\frac{\hbar}{16}\right)^3 \left(\frac{215}{64}+\frac{341 E^2}{8}+\frac{199 E^4}{4}\right)\nonumber\\
 &&+\left(\frac{\hbar}{16}\right)^4 \left(\frac{4487 E}{64}+326 E^3+\frac{1021 E^5}{4}\right)+\dots
 \label{eq:zjja}
 \end{eqnarray}
Inverting the expression for $B_{\rm ZJJ}(E, \hbar)$, we obtain \cite{Dunne:2013ada}:
\begin{eqnarray}
E_{\rm  ZJJ}(B, \hbar)&=&
B-\frac{\hbar}{16}\left(B^2+\frac{1}{4}\right)
-\left(\frac{\hbar}{16}\right)^2\left(B^3+\frac{3 B}{4}\right)
-\left(\frac{\hbar}{16}\right)^3\left(\frac{5B^4}{2}+\frac{17 B^2}{4}+\frac{9}{32}\right)\nonumber\\
   && -\left(\frac{\hbar}{16}\right)^4\left(\frac{33 B^5}{4}+\frac{205 B^3}{8}+\frac{405 B}{64}\right) -\dots
\label{bsg2}
\end{eqnarray}
which agrees with the perturbative expansion (\ref{eq:largeq}), with the definitions (\ref{hbar-g})  and  the identification of $B$ with the level number $N$:
\begin{eqnarray}
B= N+\frac{1}{2} 
\label{eq:b}
\end{eqnarray}
Thus, the function $B_{\rm ZJJ}(E, \hbar)$ is equivalent to conventional Rayleigh-Schr\"odinger perturbation theory, about the perturbative vacuum.
Using (\ref{bsg2}), the non-perturbative function $A_{\rm ZJJ}(E, \hbar)$  in (\ref{eq:zjja}) can be re-expressed as a function of $B$:
\begin{eqnarray}
 A_{\rm ZJJ}(B, \hbar)&=&\frac{16}{\hbar}+\frac{\hbar}{16}\left(3 B^2+\frac{3}{4}\right) 
 +\left(\frac{\hbar}{16}\right)^2\left(5 B^3+\frac{17 B}{4}\right)
 + \left(\frac{\hbar}{16}\right)^3 \left(\frac{55 B^4}{4}+\frac{205 B^2}{8}+\frac{135}{64}\right) \nonumber\\
&& +\left(\frac{\hbar}{16}\right)^4\frac{9}{64} \left(336 B^5+1120 B^3+327 B\right)    +\dots
\label{eq:asg}
\end{eqnarray}
The function $A_{\rm ZJJ}(B, \hbar)$ encodes the fluctuations about the single-instanton \cite{Dunne:2013ada}. For example,
the single-instanton fluctuation factor is given by
\begin{eqnarray}
\frac{\partial E_{\rm ZJJ}}{\partial B}\, e^{-\frac{1}{2} A_{\rm ZJJ}}&\sim& (1-  \frac{\hbar}{8} B - ...)\left(1-\frac{\hbar}{32}\left(3B^2+\frac{3}{4}\right)- ...\right) \nonumber\\
&=&1-\frac{\hbar}{32}\left(3B^2+4B+\frac{3}{4}\right) -\dots
\label{eq:sg-fluc}
\end{eqnarray}
in agreement with the fluctuation factor in (\ref{eq:bandwidth}). In \cite{Dunne:2013ada}, it was shown that this correspondence is directly connected with a simple relation between the two functions $A_{\rm ZJJ}(B, \hbar)$ and $E_{\rm ZJJ}(B, \hbar)$:
\begin{eqnarray}
\frac{\partial E_{\rm ZJJ}}{\partial B}=- \frac{\hbar}{16}\left(2B+\hbar\frac{\partial A_{\rm ZJJ}}{\partial \hbar}\right)
\label{eq:magic}
\end{eqnarray}
This implies that the function $A_{\rm ZJJ}(B, \hbar)$, and hence also $A_{\rm ZJJ}(E, \hbar)$,  can be deduced immediately from knowledge of the perturbative energy $E_{\rm ZJJ}(B, \hbar)$. Thus, only one of the two functions $B_{\rm ZJJ}(E, \hbar)$ and $A_{\rm ZJJ}(E, \hbar)$ is actually needed to generate the entire trans-series. Therefore, the fluctuations about the single-instanton saddle, and all other non-perturbative saddles, are precisely encoded in the fluctuations about the perturbative vacuum.  In other words, the full trans-series is encoded in the perturbative fluctuations around the vacuum, $E_{\rm ZJJ}(B, \hbar)$.  This surprising result  is in fact consistent with the ambitious goal of resurgence, which claims that  the expansion about one saddle contains, in principle, information about the expansions around other saddles, provided one knows how different saddles are connected. This connection is provided by the exact quantization condition (\ref{bloch}), which is itself a statement of the Bloch boundary condition \cite{Dunne:2013ada}.
In Section V we discuss this further, and use the gauge theory perspective to give a simple proof of this surprising result (\ref{eq:magic}).

\subsection{Electric Region: Convergent Continued-fraction Expansions, High Above the Barrier Top}
\label{sec:electric}

In the $\lambda\gg 1$ regime, which is strongly-coupling in the QM sense but weakly-coupled in the gauge theory sense, the behavior is completely different. For small level label $N$, and large $\hbar$,  the  strong-coupling expansions for low-lying modes, converted to our notation and with $u_N^{(\pm)}$ denoting the top/bottom of the $N^{\rm th}$ gap,  are (\url{http://dlmf.nist.gov/28.6.i}):
\begin{eqnarray}
u_0&=&{\hbar^2\over8}\left(0-\frac{1}{\hbar^2}+\frac{7}{4 \hbar ^6}-\frac{58}{9 \hbar ^{10}}+ \frac{68687}{2304 \hbar ^{14}}+\dots\right)\nonumber \\ 
u^{(-)}_1&=&{\hbar^2\over8}\left(1 - {4\over \hbar^2}- {2\over\hbar^4} + {1\over\hbar^6} - {1\over 6\hbar^8} - {11\over
 36 \hbar^{10}} + {49\over144 \hbar^{12}} - {55\over576 \hbar^{14}} - {83\over 540 \hbar^{16}}+\dots\right)\nonumber \\ 
u^{(+)}_1&=&{\hbar^2\over8}\left(1 + {4\over \hbar^2} - {2\over \hbar^4} - {1\over\hbar^6} - {1\over6 \hbar^8} + {11\over
 36 \hbar^{10}} + {49\over 144 \hbar^{12}} + {55\over576 \hbar^{14}} - {83\over 540 \hbar^{16}}+\dots \right)\nonumber \\ 
u^{(-)}_2&=&{\hbar^2\over8}\left(4 - {4\over 3 \hbar^4} + {5\over 54 \hbar^8} - {289\over 19440 \hbar^{12}} + {21391\over 6998400 \hbar^{16}}+\dots\right)\nonumber \\ 
u^{(+)}_2&=&{\hbar^2\over8}\left(4 + {20\over3 \hbar^4} - {763\over54 \hbar^8} + {1002401\over
 19440 \hbar^{12}} - {1669068401\over6998400 \hbar^{16}}+\dots\right)\nonumber \\ 
u^{(-)}_3&=&{\hbar^2\over8}\left(9 + {1\over\hbar^4} - {1\over\hbar^6} + {13\over80 \hbar^8} + {5\over
 16 \hbar^{10}} - {1961\over5760 \hbar^{12}} +{ 609\over6400 \hbar^{14}}+\dots\right)\nonumber \\ 
u^{(+)}_3&=&{\hbar^2\over8}\left(9 + {1\over\hbar^4} + {1\over\hbar^6} + {13\over80 \hbar^8} - {5\over
 16 \hbar^{10}} - {1961\over5760 \hbar^{12}} - {609\over6400 \hbar^{14}}+\dots\right)\nonumber \\ 
 u^{(-)}_4&=&{\hbar^2\over8}\left( 16 +{ 8\over15 \hbar^4} - {317\over3375 \hbar^8} + {80392\over
 5315625 \hbar^{12}}+\dots\right)\nonumber\\
 u^{(+)}_4&=&{\hbar^2\over8}\left( 16 +{ 8\over15 \hbar^4} +{433\over3375 \hbar^8} - {45608\over5315625 \hbar^{12}}+\dots\right)
\label{eq:conv}
\end{eqnarray}
These expansons are in fact {\it convergent}, with a radius of convergence that increases quadratically with the level index $N$.
They are generated from continued-fraction representations of the eigenvalues, and these continued-fraction expressions are themselves convergent. Nevertheless, despite these convergence properties, there are also non-perturbative effects, associated with the exponentially small splittings of the spectral gaps in this region of the spectrum, as are clearly seen in Figures \ref{fig:f1} and \ref{fig:f2}. From a physical perspective, the expansion \eqref{eq:conv} governs the ``fully quantum" regime where the kinetic term dominates over the potential, and can be obtained in a straightforward fashion by treating the potential $\cos(x)$ as a small perturbation to the free particle on a circle whose wave function is $\sim e^{i \theta x\over 2\pi }$. The even/odd wave functions are identified with $\theta=\pi N$ with $N$ being an even/odd integer. Standard degenerate perturbation theory for level  $N$ then leads to \eqref{eq:conv} where leading term is simply the energy of the particle on a circle, $N^2$, and the higher order terms are perturbative corrections to it. 

Instead of taking $\hbar\gg1$ and $N$ fixed, the high spectral region can also be probed with large $\lambda$ by taking $\hbar\to 0$ and $N\to\infty$, with $N\hbar\gg 1$.
Then for large level number $N\gg 1/\hbar$,  the continued-fraction expressions for the energy eigenvalues give  approximate expressions for the energy of the $N^{\rm th}$  gap as (\url{http://dlmf.nist.gov/28.6.E14}):
 \begin{eqnarray}
u(N, \hbar) 
& \sim &
\frac{\hbar^2}{8}\left(N^2+\frac{1}{2(N^2-1)}\left(\frac{2}{\hbar}\right)^4
+\frac{5N^2+7}{32(N^2-1)^3(N^2-4)} \left(\frac{2}{\hbar}\right)^8 \right. \nonumber\\
&&\left .+\frac{9N^4+58 N^2+29}{64(N^2-1)^5(N^2-4)(N^2-9)}\left(\frac{2}{\hbar}\right)^{12}+\dots\right)
\label{eq:smallq}
\end{eqnarray}
The continued fraction relation that generates the above expansion is given in Section \ref{sec:strong-weak}. Note that each coefficient has poles at integer values of $N$. In particular, the denominator of the coefficient of $\hbar^{2-4n}$ is proportional to 
\bea
\prod_{k=1}^{n} (N^2-k^2)^{2 \left \lfloor{n\over k}\right \rfloor-1}
\ea
where $\left \lfloor{n}\right \rfloor$ denotes the greatest integer less than or equal to $n$. 

Now we demonstrate how the energy spectrum \eqref{eq:smallq} is identified with the multi-instanton expansion of the prepotential (\ref{eq:ns-prepotential}) in the Nekrasov-Shatashvili limit of the $\mathcal N=2$ $SU(2)$ SUSY gauge theory. First, rewrite (\ref{eq:smallq}) as:
\begin{eqnarray}
u&\sim& {1\over 2}\left(\frac{N\hbar}{2}\right)^2+ \frac{1}{4} \left(\frac{2}{N\hbar}\right)^2 \frac{1}{\left(1-\frac{\hbar^2}{(N \hbar)^2}\right)}+\frac{5}{64} \left(\frac{2}{N\hbar}\right)^6\frac{\left(1+\frac{7\hbar^2}{5(N \hbar)^2}\right)}{\left(1-\frac{\hbar^2}{(N \hbar)^2}\right)^3\left(1-\frac{4\hbar^2}{(N \hbar)^2}\right)}+\dots 
 \\
&\sim& \left[{a^2\over2}+ \frac{1}{4\,a^2}+\frac{5}{64} \frac{1}{a^6}+\frac{9}{128} \frac{1}{a^{10}}+\dots\right]
+\hbar^2\left[\frac{1}{16\,a^4}+\frac{21}{128} \frac{1}{a^8}+\frac{55}{128} \frac{1}{a^{12}}+\dots \right]
+\dots
\label{eq:smallq-u}
\end{eqnarray}
where we have defined the ``action''
\begin{eqnarray}
a\equiv \frac{N\, \hbar}{2} 
\label{eq:action}
\end{eqnarray}
which is half the ``'t Hooft coupling'' defined previously (\ref{thooft}). We now compare this with the instanton expansion in \eqref{eq:ns-prepotential}. In order to relate the prepotential to $u$ we use Matone's relation \cite{Matone:1995rx,Bilal:1996sk,Poghossian:2010pn}, 
\bea
u(a,\hbar)={i\pi\over2}\Lambda {\partial \mc F_{NS}\over \partial \Lambda} -{\hbar^2 \over 48}\,. 
\label{eq:Matone}
\ea
The second term on the right hand side is due to the perturbative part of the prepotential.  Plugging the expansion 
\bea
\mathcal F_{NS}=&&\left( -{a^2\over 2\pi i}\log{a^2\over\Lambda^2}+{\Lambda^4\over 8 \pi i a^2} + {5\Lambda^8\over256 \pi i a^6} +\dots\right) -\frac{\hbar^2}{48 \pi i}\log\frac{a ^2}{\Lambda ^2}+{\hbar ^2\over2 \pi i} \left(\frac{\Lambda ^4}{16 a ^4}+\frac{21 \Lambda ^8}{256 a ^8}+\dots\right)\nonumber\\&&+\dots 
\ea
into \eqref{eq:Matone} one sees that the expansion for $u(a,\hbar)$ matches precisely with the expansion in the second line of \eqref{eq:smallq-u}.
 
However, it is clear that (\ref{eq:smallq}, \ref{eq:smallq-u}) is not the whole story. First, this gives the same expression for both gap edges, $u_N^{(\pm)}$, whereas the expressions (\ref{eq:conv}) clearly show that $u_N^{(\pm)}$ are different from one another. Second, the $\hbar$ dependence of the expansion (\ref{eq:smallq}) does not match that of (\ref{eq:conv}). Physically, this is simply the statement that the perturbation theory 
Bohr-Sommerfeld expansion (\ref{eq:smallq}) is only approximate, giving the approximate {\it location} of a narrow gap high in the spectrum (see Figures \ref{fig:f1} and \ref{fig:f2}) for $N\gg 1/\hbar$, completely neglecting the non-perturbative splitting for the {\it width} of the gap. 
In fact, for a given $N$, the splitting of the energy levels occurs at the order $1/\hbar^{2N}$. As the level index $N$ increases, the splitting drifts to higher orders in perturbation theory and becomes exponentially small
(\url{http://dlmf.nist.gov/28.6.E15}):
\begin{eqnarray}
\Delta u^{\rm gap}_N
&\sim& \frac{\hbar^2}{4} \frac{1}{\left(2^{N-1} (N-1)! \right)^2} \left(\frac{2}{\hbar}\right)^{2N}  \left[1+O\left(\left(\frac{2}{\hbar}\right)^4\right) \right]  \nonumber \\
&\sim & \frac{N\, \hbar^2}{2\pi}\left(\frac{e}{N\, \hbar}\right)^{2N}\qquad, \quad N\gg 1
\label{eq:smallq-splitting}
\end{eqnarray}
Since the expansions (\ref{eq:conv}) are convergent, the origin of this non-perturbative splitting is quite different from the familiar weak-coupling analysis that associates non-perturbative terms with divergent perturbative series \cite{LeGuillou:1990nq,zjbook,kenbook}. It is clear from the continued-fraction expansion  (\ref{eq:smallq}) that the splitting is directly associated with the  poles of the expansion coefficients at integer values of $N$.  Physically this indicates the appearance of degenerate perturbation theory for both edges of the gap \cite{peierls}. 
As mentioned already in the Introduction, this is a concrete analog of the results Drukker, Mari\~no and Putrov \cite{Drukker:2010nc} concerning the large $N$ expansion of the ABJM matrix model, in which non-perturbative effects are related to complex space-time instantons, and which were subsequently related to poles in the 't Hooft expansion coefficients \cite{Hatsuda:2012hm}.
We show in Section IV that in this regime the gap splitting (\ref{eq:smallq-splitting}) also has a natural non-perturbative interpretation in terms of semiclassical configurations. 

To make a sharper analogy with the trans-series (multi-instanton) expansion \eqref{eq:trans} near the bottom of the well, we can reorganize the large $\hbar$ expansion \eqref{eq:conv} as
\bea
u^{(\pm)}_N(\hbar)=\frac{\hbar^2  N^2}{8}\sum_{k=0}^\infty  \frac{\alpha_k(N)}{\hbar^{4k}} \pm \frac{\hbar^2}{8}{1\over (2^{N-1}(N-1)!)^2} \left(\frac{2}{\hbar}\right)^{2N} \sum_{k=0}^\infty \frac{\beta_k(N)}{\hbar^{4k}} 
\label{eq:smallq-alt}
\ea
where the coefficients $\alpha_k(N)$ are those given in the continued fraction expansion \eqref{eq:smallq}. However this identification is valid only up to order $k=(N-1)/2$, due to the poles in the coefficients of the continued fraction expansion. The gap splitting, encoded in the $\beta_k(N)$ terms, arises beyond this order, and has leading behavior  given in \eqref{eq:smallq-splitting}. So, at finite $N$ the gap splitting terms are missed by the Bohr-Sommerfeld expansion.

In the next Section we show how we can actually obtain these non-perturbative gap splittings from an exact all-orders WKB analysis. In Section \ref{sec:wli} we will further identify this effect with worldline instantons, and make a direct physical analogy with multi-photon ionization in monochromatic time-dependent laser pulses.

\section{All-orders WKB Analysis of the Mathieu Equation: Actions and Dual Actions}
\label{sec:wkb}

\begin{figure}[htb]
\includegraphics[scale=0.3]{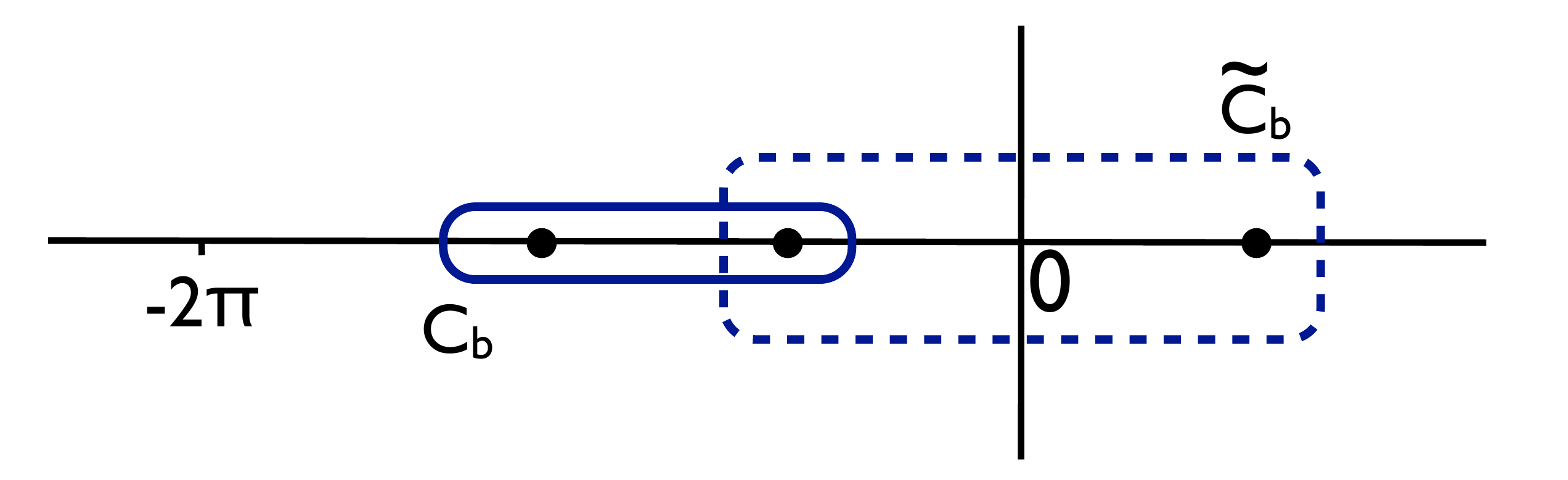}
\includegraphics[scale=0.3]{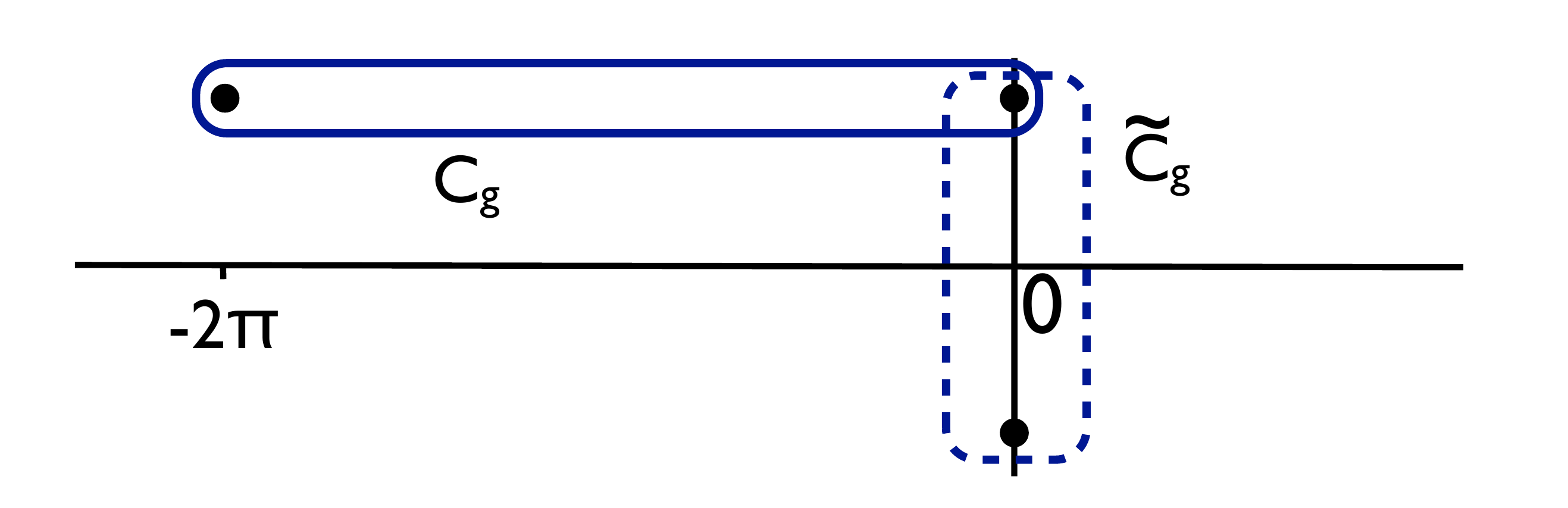}
\caption{The integration contours for the all-orders WKB expressions. The first figure shows the paths for the bands, when the energy is well below the barrier top. The contour $C_b$ is used to determine the location of the band, while $\tilde{C}_b$ is used to determine the width of the band. The second figure shows the paths for the gaps, when the energy is well above the barrier top. The contour $C_g$ is used to determine the location of the gap, while $\tilde{C}_g$ is used to determine the width of the gap. Note that in this case the relevant turning points are in the complex plane.}
\label{fig:f4}
\end{figure}

\begin{figure}[htb]
\center
\includegraphics[scale=0.3]{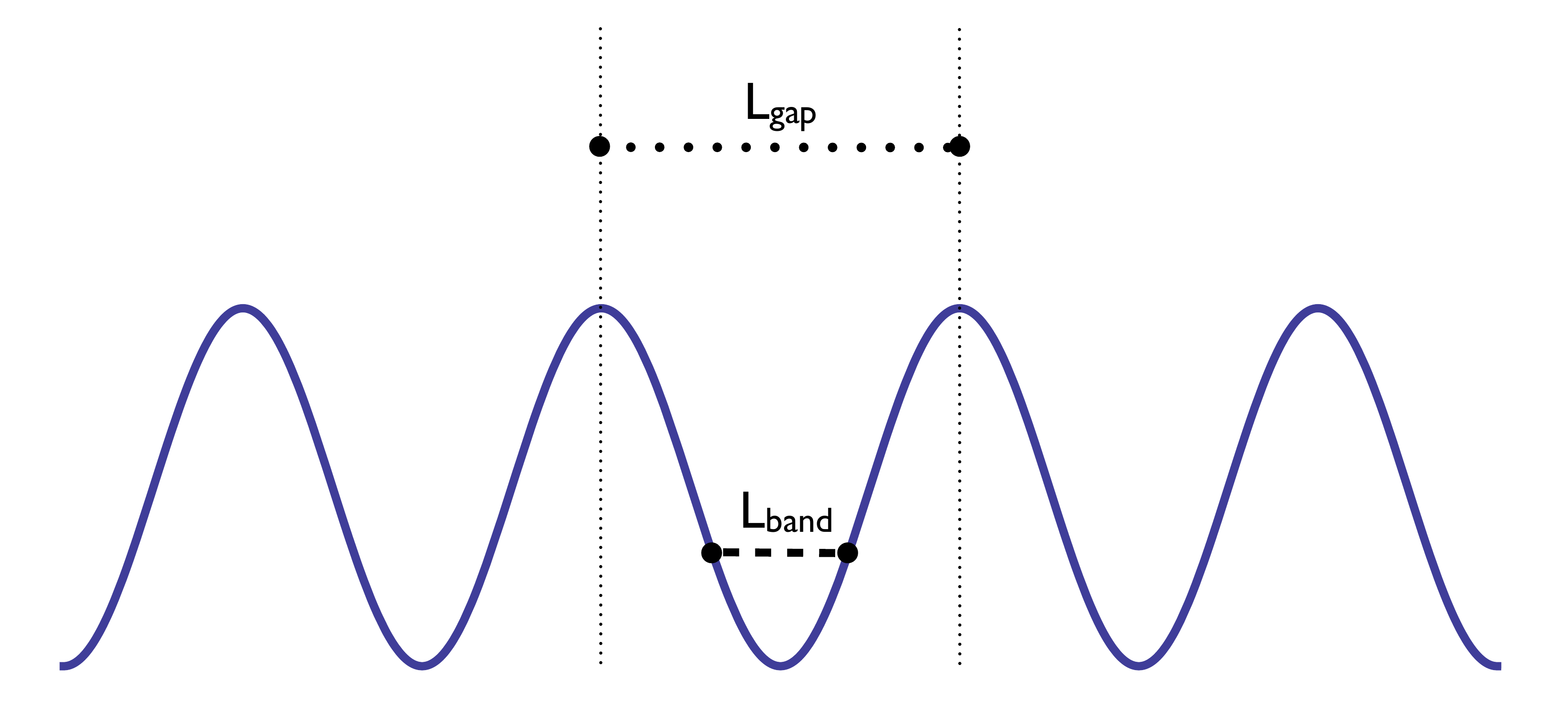}
\caption{To determine the {\it location} of the bands or gaps, we integrate the WKB integrands along the paths, $L_{\rm band}$ and $L_{\rm gap}$, shown here. In the deep band case, the {\it width} of the band is determined by integrating through the barrier. As the energy goes above the barrier top, the turning points coalesce and move into the complex plane: in the high gap case, the {\it width} of the gap is determined by integrating between these complex turning points.}
\label{fig:f5}
\end{figure}

The all-orders WKB expression for the location of bands and gaps can be expressed as \cite{dunham,bender,voros,takei,delabaere,howls-book}:
\begin{eqnarray}
\oint_C P =
\begin{cases}
2 \pi N\, \hbar \qquad\qquad \quad ({\rm gap}) \cr
2 \pi \left(N+\frac{1}{2}\right)\hbar \qquad ({\rm band})
\end{cases}
\end{eqnarray}
where $P$ is the local momentum and the contours go around the appropriate  turning points, as shown in Figure \ref{fig:f4}. Geometrically, for the Mathieu problem there are two independent cycles and they correspond to the generators of the two cycles of the torus.
More explicitly, with $u$ denoting the energy eigenvalue, we have \cite{dunham,bender}
\begin{eqnarray}
{1\over 2\pi} \oint P &=& \frac{\sqrt{2}}{2\pi}\left(\oint_C \sqrt{u-V} dx-\frac{\hbar^2}{2^6}\oint_C \frac{(V^\prime)^2}{(u-V)^{5/2}} dx-\frac{\hbar^4}{2^{13}}\oint_C \left(\frac{49 (V^\prime)^4}{(u-V)^{11/2}}- \frac{16 V^\prime V^{\prime\prime\prime}}{(u-V)^{7/2}}\right)dx-\dots \right)\nonumber\\
&=&\begin{cases}
N\, \hbar \qquad\quad \quad ({\rm gap}) \cr
\left(N+\frac{1}{2}\right)\hbar \quad ({\rm band})
\end{cases}
\end{eqnarray}
With proper analytic continuation, this quantization condition permits smooth transitions and dualities between the various spectral regions, connecting weak and strong coupling, and also the bottom and top of the wells. The distinction between the various regions is encoded in the location of the turning points in the complex plane, and the associated Stokes lines \cite{delabaere}.  For energies inside the wells there are real turning points. As the energy approaches the barrier top,  the turning points  come together and coalesce, and move apart again along the imaginary axis for energy above the barrier top.
See Figures \ref{fig:f4} and \ref{fig:f5} .

For example, the leading behavior in (\ref{eq:largeq}) and (\ref{eq:smallq}) can be found immediately as follows. For the location of the center of a high gap:
\begin{eqnarray}
N \hbar \approx \frac{2\sqrt{2}}{\pi}  \int_{-1}^1 \sqrt{\frac{u-y}{1-y^2}}\, dy  \sim 2\sqrt{2u} +O\left(\frac{1}{u^{3/2}}\right) 
\quad \Rightarrow\quad u(N, \hbar)\sim \frac{N^2\hbar^2}{8}+\dots
\end{eqnarray}
For the location of the center of a deep band:
\begin{eqnarray}
\left(N+\frac{1}{2}\right)\hbar \approx \frac{2\sqrt{2}}{\pi}  \int_{-1}^u \sqrt{\frac{u-y}{1-y^2}}\, dy  \sim (1+u)+O\left((1+u)^2\right) 
\,\Rightarrow\, u(N, \hbar)\sim -1+\hbar\left(N+\frac{1}{2}\right)+\dots\nonumber\\
\end{eqnarray}
The higher terms are discussed below, in the next subsections.

Perhaps less well-known is that there are corresponding expressions for the {\it width} of a band deep in the spectrum  and of a gap high in the spectrum \cite{landau,froman,dykhne,connor-marcus,weinstein-keller}. At leading order, the width of a band or a gap is expressed as the product of a density-of-states factor and an exponential:
\begin{eqnarray}
\Delta u_N &\sim& \frac{2}{\pi} \frac{\partial u}{\partial N} \exp\left[-\frac{1}{\hbar} {\rm Im}\, \oint_{\tilde{C}} P\right]
\end{eqnarray}
where now the contours are around the dual contours $\tilde{C}_b$ or  $\tilde{C}_g$ shown in Figure \ref{fig:f4}. Leading order results for bands and gaps in the barrier region are discussed in \cite{Neuberger:1978ft,froman,weinstein-keller,connor-marcus}, and in more detail  below in Section \ref{sec:wkb-magnetic}.

As discussed in the introduction, the WKB periods associated with the Mathieu and Lam\'e equations correspond to the vacuum expectation value of the scalar field $\Phi$ and its dual partner, and the energy eigenvalue $u$ correspond to the gauge invariant observable $\langle{\rm Tr}\, \Phi^2\rangle$. We set our normalization such that classically $\Phi_{cl}(a)={a\over 2}\sigma^3$, where $\sigma^3$ is the $3^{rd}$ Pauli matrix,  and therefore $a=\sqrt{2u}+\dots$, fixing the normalization to
\bea
a(u)&:=&{1\over 4 \pi}\oint_{\gamma_1} P= {\sqrt{2}\over 2\pi}\left(\int_{-\pi}^\pi\sqrt{u-V}dx-\frac{\hbar^2}{2^6}\int_{-\pi}^\pi \frac{(V^\prime)^2}{(u-V)^{5/2}}dx -\dots\right)\\
 a_D(u)&:=&{1\over 4 \pi}\oint_{\gamma_2} P= {\sqrt{2}\over 2\pi}\left(\int_{-\cos^{-1}(u)}^{\cos^{-1}(u)}\sqrt{u-V}dx-\frac{\hbar^2}{2^6} \int_{-\cos^{-1}(u)}^{\cos^{-1}(u)}\frac{(V^\prime)^2}{(u-V)^{5/2}}dx -\dots\right)
 \label{eq:BSperiods}
\ea

\subsection{Dyonic Region: Resurgence from All-Orders WKB}
\label{sec:wkb-dyonic}

For energies below the barrier top, where $-1\leq u\leq 1$, the leading WKB expression for the band location is given by the usual Bohr-Sommerfeld expression that involves only the real part of the action, while the dual action is associated with under-the-barrier, and so is pure imaginary\footnote{Here, and for the rest of the paper, the notion of ``real''/``pure imaginary'' action refers to the particular choice of cycles such that when $-1\leq u < \infty $, and $\hbar \in \mathbb R^+$, the associated action is real/pure imaginary. In general, $u$ and $\hbar$ can take complex values, and the analysis follows by analytic continuation.}. The real action in this region can be expressed as a linear combination of the two independent actions defined above:
\bea
\mc Re[a(u)]=a(u)+a^D(u)
\ea
The exact quantization that identifies $u$ with the center of the band is implemented by requiring
\begin{eqnarray}
Re\left[a(u, \hbar)\right]= a(u)+a^D(u)=\frac{\hbar}{2}\,  \left(N+\frac{1}{2}\right)\,.
\label{eq:bs}
\end{eqnarray}
The leading order terms of the actions are expressed in terms of elliptic integrals
\begin{eqnarray}
\mc Re[a_0(u)]&=&{\sqrt{2}\over\pi}\, \int_{-1}^1 dy\, \sqrt{\frac{y-u}{y^2-1}}  
= {2\over \pi} \left({\mathbb E}\left(\frac{u+1}{2}\right)-\frac{1}{2} (1-u)\, {\mathbb K}\left(\frac{u+1}{2}\right)\right)\\
a_0^D(u)&=&- {i\sqrt{2}\over\pi}\, \int_{u}^1 dy\, \sqrt{\frac{y-u}{y^2-1}}  
= -{2i\over \pi} \left({\mathbb E}\left(\frac{1-u}{2}\right)-\frac{1}{2} (u+1)\, {\mathbb K}\left(\frac{1-u}{2}\right)\right)\,.
\label{aperiods}
\end{eqnarray}
These expressions are the original Seiberg-Witten solution, which in the quantum mechanical language is identified with the leading order WKB expansion. Note that they satisfy the Picard-Fuchs relation:
\begin{eqnarray}
a_0(u) \frac{d a_0^D(u)}{du}- a_0^D(u) \frac{d a_0(u)}{du} ={2 i\over \pi} 
\label{eq:picard}
\end{eqnarray}
which follows from the Legendre relation [here ${\mathbb K}^\prime$ denotes ${\mathbb K}(1-k^2)$, using the conventions of \cite{nist}]:
\begin{eqnarray}
{\mathbb E}{\mathbb K}' +{\mathbb E}' {\mathbb K} -{\mathbb K} {\mathbb K}'=\frac{\pi}{2}\,.
\label{eq:legendre}
\end{eqnarray}
The Picard-Fuchs equation \eqref{eq:picard} is invariant under $SL(2,\mathbb Z)$ transformations applied to the pair $(a_0,a^D_0)$ which corresponds to changing the basis for the two cycles \cite{Klemm:1995wp,Matone:1995rx,Bilal:1996sk,Poghossian:2010pn}. For example, we can replace $a_0$ by $\mc \mc Re[a_0]$ in \eqref{eq:picard} and the equation equation will still hold.  

Remarkably, the higher-order WKB actions can be obtained by acting on these leading WKB actions with  differential operators with respect to the energy $u$ \cite{Mironov:2009uv,He:2010xa}. This follows from the fact that for $V=\cos(x)$, the numerators in \eqref{eq:BSperiods}, which are given by the derivatives of $V$, can be re-expressed as polynomials of $V$. Therefore by differentiating $\sqrt{u-V}$ with respect to $u$ taking appropriate combinations one can generate the integrands in \eqref{eq:BSperiods} up to total derivatives which vanish after integrating over turning points. For example, at the next two orders:
\bea
a_1(u)&=&{1\over 48}\left(2u{d^2\over du^2}+{d\over du}\right) a_0(u)\\
a_2(u)&=&{1\over 2^9 45}\left(28 u^2{d^4\over du^4}+120u{d^3\over du^3}+75{d^2\over du^2}\right) a_0(u)
\label{eq:ans}
\ea
and the same relations hold for $a^D(u)$ as well.  The general form of the differential operators that relates $a_n$ to $a_0$ is
  \bea
  a_n(u)=\sum_{k=0}^n \kappa^{(n)}_k u^k {d^{n+k} a_0(u)\over du^{n+k}}\,.
  \label{eq:ans-general}
  \ea
  We have verified \eqref{eq:ans-general} to the order $\hbar^{10}$. The coefficients $\kappa^{(n)}_k$ up to this order are given in Appendix \ref{app:kappas}. The first two next-to-leading order actions calculated from \eqref{eq:ans} are 
     \begin{eqnarray}
\mc Re[a_1(u)]&=& \frac{1}{48 \pi  \left(1-u^2\right)}\left((1-u) {\mathbb K}\left(\frac{1+u}{2}\right)+2 u \, {\mathbb E}\left(\frac{1+u}{2}\right)\right)\\
   a_1^D(u)&=& \frac{i}{48 \pi  \left(1-u^2\right)}\left( (1+u)\,  {\mathbb K}\left(\frac{1-u}{2}\right) -2 u\,  {\mathbb E}\left(\frac{1-u}{2}\right)\right)   
   \end{eqnarray}
      \begin{eqnarray}
\mc Re[a_2(u)]=-\frac{1}{46080 \pi \left(1-u^2\right)^3}&&\left[ (1-u) (4u^3+93 u^2-60 u+75) {\mathbb K}\left(\frac{1+u}{2}\right) \right. \nonumber  \\ 
                 &&\left.+2 \left(4 u^4-153 u^2-75\right) {\mathbb E}\left(\frac{1+u}{2}\right)\right] \\ 
   a_2^D(u)=\frac{i}{46080 \pi \left(1-u^2\right)^3}&&\left[(1+u) (-4u^3+93 u^2-60u^2+75) {\mathbb K}\left(\frac{1-u}{2}\right) \right. \nonumber \\
   &&\left.+2\left(4 u^4-153 u^2-75\right) {\mathbb E}\left(\frac{1-u}{2}\right)\right]
   \end{eqnarray}
The actions can be expanded about the bottom of the wells, $u\sim -1$, as:
\begin{eqnarray}
Re\left[a_0(u)\right] &\sim& \frac{u+1}{2}+\frac{ (u+1)^2}{32} +\frac{3  (u+1)^3 }{512}
+\frac{25 (u+1)^4}{16384}+
 \frac{245 (u+1)^5}{524288}+\dots\\\
Re\left[a_1(u)\right]&\sim&\frac{1}{128}+\frac{5(u+1)}{2048}+ \frac{35 (u+1)^2}{32768} +
\frac{525 (u+1)^3}{1048576}+
\frac{8085 (u+1)^4}{33554432}+\dots \\
Re\left[a_2(u)\right]&\sim& \frac{17}{262144} + \frac{721 (u+1)}{8388608}+ \frac{10941(u+1)^2}{134217728}+ \frac{141757 (u+1)^3}{2147483648}
+\dots
\end{eqnarray}
Recalling (\ref{hbar-g}) that $1+u\equiv \hbar E$, we can re-write these expressions for $\mc Re[a_n(u)]$ as expansions in powers of $\hbar$ and $E$:
\begin{eqnarray}
Re\left[\frac{2}{ \hbar} \,a_0(u)\right] &\sim& E+\left(\frac{\hbar}{16}\right) E^2+\left(\frac{\hbar}{16}\right)^2 3 E^3 
+\left(\frac{\hbar}{16}\right)^3 \frac{25}{2} E^4+\left(\frac{\hbar}{16}\right)^4 \frac{245}{4} E^5 +\dots \\
\hbar^2 \, Re\left[\frac{2}{ \hbar} \,a_1(u)\right] &\sim& \frac{1}{4}\left(\frac{\hbar}{16}\right)+\left(\frac{\hbar}{16}\right)^2 \frac{5}{4} E+\left(\frac{\hbar}{16}\right)^3 \frac{35}{4} E^2 + \left(\frac{\hbar}{16}\right)^4 \frac{525}{8} E^3+\dots\\
\hbar^4 \, Re\left[\frac{2}{ \hbar}\, a_2(u)\right] &\sim& \frac{17}{32}\left(\frac{\hbar}{16}\right)^3+\left(\frac{\hbar}{16}\right)^4 \frac{721}{64} E
+\dots
\end{eqnarray}
Comparing with (\ref{eq:zjjb}), 
we recognize these expansions as the highest powers of $E$, the next-to-highest powers of $E$, and next-to-next-to-highest powers of $E$, respectively, for each power of $\hbar$ in  the function $B_{ZJJ}(E, \hbar)$. Thus we have the following identification with the results of Zinn-Justin and Jentschura \cite{zjj}:
\begin{eqnarray}
B_{ZJJ}(E, \hbar) &=& \frac{2}{\hbar} \sum_{n=0}^\infty \hbar^{2n} \, Re\left[a_n(-1+\hbar E)\right] \\
&\equiv & \frac{2}{\hbar} \, Re\left[a(-1+\hbar E, \hbar) \right]
\label{eq:b2}
\end{eqnarray}
Thus the ``exact Bohr-Sommerfeld condition'' \eqref{eq:bs} for the location of the energy bands deep inside the wells expresses the perturbative expansion for the location of the $N^{\rm th}$ band. Inverting this Bohr-Sommerfeld condition, to express $u=u(N, \hbar)$ as a function of $N$ and $\hbar$, we arrive at the perturbative expansion (\ref{eq:largeq}), or in the notation of ZJJ, the expression  (\ref{bsg2}) for the energy $E(B, \hbar)$ as an expansion in $\hbar$ and $B\equiv N+\frac{1}{2}$. Recall that this expansion  (\ref{bsg2}) is non-Borel-summable.

Similarly, the dual actions can be expanded about the bottom of the wells, $u\sim -1$, as:
\begin{eqnarray}
-i\, a_0^D(u) &\sim& -{4\over \pi}-\frac{(u+1)}{2\pi} \left(\log \left(\frac{u+1}{32}\right)-1\right)-
\frac{ (u+1)^2}{64\pi} \left(2 \log \left(\frac{u+1}{32}\right)+3\right)+\dots \\
-i\, a_1^D(u) &\sim& \frac{1}{48 \pi (u+1)}-\frac{1}{384 \pi } \left(3 \log  \left( \frac{u+1}{32}\right)+5\right)
- \frac{(u+1)}{12288 \pi } \left(30 \log
  \left( \frac{u+1}{32}\right) +77\right) \nonumber \\
   &&-\frac{(u+1)^2}{32768\pi} \left((35\log \left( \frac{u+1}{32}\right)+94\right)  -\dots \\
-i\, a_2^D(u) &\sim& -\frac{7}{5760 \pi (u+1)^3}  -\frac{1}{10240 \pi (u+1)^2} -\frac{53}{491520 \pi (u+1)} \nonumber \\
&&-\frac{1}{23592960\pi} \left(1530 \log \left( \frac{u+1}{32}\right)+6227\right)  \nonumber \\
&&-
\frac{(u+1)}{503316480\pi}  \left(43260 \log \left( \frac{u+1}{32}\right) +152759\right)  -\dots 
\end{eqnarray}
Notice that for $n\geq 1$, the $a_n^D(u)$ diverge as $u\to -1$.

The dual action $a^D(u, \hbar)$ is related to Zinn-Justin's $A_{ZJJ}(E, g)$ function as follows:
\begin{eqnarray}
A_{ZJJ}(E, \hbar)&=&\frac{4\pi i}{\hbar} \,a^D(-1+\hbar\, E)-2\ln \Gamma\left(\frac{1}{2}+B(E, \hbar)\right)+\ln(2\pi)-2B(E, \hbar)\, \ln\left(\frac{\hbar}{32}\right) \\
&=&-\frac{4 \pi }{\hbar} \,Im\left(a^D(-1+\hbar\, E)\right)-2\ln \Gamma\left(\frac{1}{2}+B(E, \hbar)\right)+\ln(2\pi)-2B(E, \hbar)\, \ln\left(\frac{\hbar}{32}\right)\nonumber \\
\label{a}
\end{eqnarray}
The comparison for $A(E, \hbar)$ is non-trivial, as it requires using the large $B$ asymptotics of the $\ln\Gamma$ function. The subtraction of these terms corresponds to matching the perturbative solution near the bottom of the well to the solution near the top of the barrier, coming from the exact solution for an inverted harmonic well, which is how $a^D(u)$ looks near the top of the barrier \cite{froman,connor-marcus}.

\subsection{Electric Region: Convergent Expansions and the Nekrasov Instanton Expansion}
\label{sec:wkb-electric}
In the electric region (i.e. $u\rightarrow \infty$) the real period is identified with $a(u)$ and the pure imaginary period is still $a^D(u)$.  The exact quantization that identifies $u$ with the center of the gap is 
\bea
a(u)={\hbar\over2} N 
\label{eq:electric-quant}
\ea
The actions can be expanded for $u\gg 1$ as:
\begin{eqnarray}
a_0(u)&\sim& \sqrt{2u}\left( 1-\frac{1}{16 u^2}-\frac{15}{1024 u^4} -\frac{105}{16384 u^6}-\dots 
\right) \\
a_1(u)&\sim&-\frac{1}{16 \left(2u\right)^{5/2}} \left(1+\frac{35}{32 u^2}+\frac{1155}{1024 u^4}
+\frac{75075}{65536 u^6}+\dots\right)  \\
a_2(u)&\sim&-\frac{1}{64 \left(2u\right)^{7/2}}\left(1+ \frac{273}{64 u^2} +\frac{5005}{512 u^4}+
\frac{2297295}{131072 u^6}+\dots\right)  
\end{eqnarray}
Combining these expansions we find
\begin{eqnarray}
\frac{2}{ \hbar}\left(a_0(u)+\hbar^2 a_1(u) +\hbar^4 a_2(u)+\dots\right)
&\sim& \frac{2\sqrt{2 u}}{\hbar} \left[1-\frac{1}{16 u^2}-\frac{15}{1024 u^4}-\frac{105}{16384 u^6} 
-\dots\right] \nonumber\\
&& -\frac{\hbar}{8 (2 u)^{5/2}}\left[ 1+\frac{35}{32 u^2}+\frac{1155}{1024 u^4}
+\frac{75075}{65536 u^6}
+\dots \right]  \nonumber \\
&&- \frac{\hbar^3}{32 (2 u)^{7/2}} \left[1+ \frac{273}{64 u^2} +\frac{5005}{512 u^4}+
\frac{2297295}{131072 u^6}
+ \dots\right]-\dots \nonumber \\
\label{eq:electric-wkb-expansion}
\end{eqnarray}
Identifying the left-hand-side with $N$,  and inverting, we obtain the  expansion (\ref{eq:smallq-u}) of the gap energy $u(N, \hbar)$. 
Thus, the all-orders-WKB action $a(u, \hbar)$ determines the (convergent) expansion of the location of the gap high up in the spectrum.

The dual actions can be expanded for $u\gg 1$ as:
\begin{eqnarray}
-i a_0^D(u) &\sim& {\sqrt{2u}\over \pi}\left(-2+\log (8 u)+\frac{1-\log (8 u)}{16 u^2} + 
\frac{47-30 \log (8u)}{2048 u^4}  +\dots \right)\\
-ia_1^D(u) &\sim& \frac{1}{24\pi\sqrt{2u}}\left(1+\frac{13-6 \log (8u)}{16
   u^2} + \frac{883-420 \log (8u)}{1024 u^4}+\dots \right)\\
 -ia_2^D(u) &\sim& \frac{1}{45 2^6 \pi(2u)^{3/2}}\left(-1+\frac{567-180 \log (8 u)}{16 u^2}+
     \frac{127461-49140 \log (8 u)}{1024 u^4}+\dots \right)
\end{eqnarray}
These dual actions determine the exponentially narrow width of the gap, high up in the spectrum, as discussed in the next subsection.

\subsection{Magnetic Region: Duality and Analytic Continuation Across the Barrier}
\label{sec:wkb-magnetic}

Across  the magnetic region, there is a transition from the divergent perturbative behavior characteristic of the dyonic region, and the convergent perturbative expansions characteristic of the electric region. As is clear from the plots in Figures \ref{fig:f1} and \ref{fig:f2}, the transition is smooth, but connecting the regions requires careful interpretation of the various expansions. Of particular interest are the different mechanisms by which non-perturbative terms arise in the different physical regions. For example, the general expression for the exponentially narrow width of a band deep in the dyonic region, and of a gap high in the electric region is \cite{froman,weinstein-keller,connor-marcus}: 
\begin{eqnarray}
\Delta u\sim \frac{2}{\pi}\frac{\partial u}{\partial N}\, e^{-\frac{2 \pi}{\hbar} {\rm Im}\, a_0^D}
\sim {\hbar\over \pi} \frac{\partial u}{\partial \mc Re[a_0]}\, e^{-\frac{2 \pi}{\hbar} {\rm Im}\, a_0^D} 
\label{eq:bandgap}
\end{eqnarray}
In the dyonic region, 
\begin{eqnarray}
u&\sim & -1+2\,\mc Re[a_0(u)] +\dots = -1+\hbar\left(N+\frac{1}{2}\right) +\dots \\
\pi \,\mc Im [ a_0^D] &\sim& 4+\frac{1+u}{2}\left(\ln\left(\frac{1+u}{32}\right) -1\right)+\dots 
\label{eq:dyonic-limits}
\end{eqnarray}
Therefore, from (\ref{eq:bandgap}) we obtain the band width estimate (using Stirling's formula in the last step):
\begin{eqnarray}
\Delta u^{\rm band} &\sim& \frac{2\hbar}{\pi} \left(\frac{\hbar\left(N+\frac{1}{2}\right)}{32\, e}\right)^{-(N+\frac{1}{2})} \, e^{-8/\hbar} \nonumber\\
&\sim& \sqrt{\frac{2}{\pi}}\frac{2^{4(N+1)}}{N!} \left(\frac{2}{\hbar}\right)^{N-\frac{1}{2}}\, e^{-8/\hbar}
\label{eq:dyonic-band-width}
\end{eqnarray}
in agreement with (\ref{eq:bandwidth}).  On the other hand, in the electric region,
\begin{eqnarray}
u&\sim & \frac{1}{2} a_0^2 +\dots  = \frac{\hbar^2}{8} N^2 +\dots \\
\pi \,\mc Im[ a_0^D] &\sim& \sqrt{2u}\left(\ln(8 u)-2\right)+\dots 
\label{eq:electric-limits}
\end{eqnarray}
Therefore, from (\ref{eq:bandgap}) we obtain the gap width estimate:
\begin{eqnarray}
\Delta u^{\rm gap} &\sim& \frac{\hbar^2 N}{2\pi} \left(\frac{e}{\hbar N}\right)^{2N} 
\label{eq:electric-gap-width}
\end{eqnarray}
in agreement with (\ref{eq:smallq-splitting}). Thus, the formula (\ref{eq:bandgap}) has the correct form in both extreme limits, in one case referring to the width of a band, and in the other to the width of a gap.

The magnetic region near $u\sim 1$ is more subtle. In this regime,
\begin{eqnarray}
a_0&\sim& {4\over \pi} +\frac{u-1}{2\pi}\left[\ln\left(\frac{32}{u-1}\right) +1\right]+\dots \\
-i a_0^D&\sim& \frac{1}{2}\left(u-1\right)+\dots
\label{eq:barrier-limits}
\end{eqnarray}
The latter relation tells us that the exponential behavior becomes of order unity. The first relation gives us the leading scaling between $N$ and $\hbar$:
\begin{eqnarray}
N\sim \frac{8}{\pi\,\hbar}
\label{eq:top}
\end{eqnarray}
It is clear that the barrier top is in the vicinity of  $N \sim 1/\hbar$, but the above fixes the non-trivial coefficient to be $8/\pi$. It is instructive to evaluate the energy eigenvalue $u(N, \hbar)$ with this scaling, in both the dyonic and electric regions, using (\ref{eq:largeq}) and (\ref{eq:smallq}) respectively:
\begin{eqnarray}
u_{\rm dyonic}&\sim& -1+\frac{8}{\pi}\left[1-\frac{1}{16}\frac{8}{\pi}-\frac{1}{2^8}\left(\frac{8}{\pi}\right)^2-\frac{5}{2^{14}}\left(\frac{8}{\pi}\right)^3 -\frac{33}{2^{18}}\left(\frac{8}{\pi}\right)^4-\dots \right] +O(\hbar)\nonumber\\
&=& 1+O(\hbar)  \\
u_{\rm electric}&\sim& \frac{1}{2} \left[\left(\frac{4}{\pi}\right)^2 + \frac{1}{2} \left(\frac{\pi}{4}\right)^2+\frac{5}{32} \left(\frac{\pi}{4}\right)^6+\frac{9}{64}\left(\frac{\pi}{4}\right)^{10}+\dots \right]+O(\hbar) \nonumber\\
&=&1+O(\hbar)
\label{eq:barriertop}
\end{eqnarray}
It is remarkable that these two very different expansions coincide at $u\sim 1$.

In fact, we can refine further the estimate in (\ref{eq:top}). The {\it edges} of the bands/gaps when $u=1$ are given by \cite{weinstein-keller}
\begin{eqnarray}
N\pm \frac{1}{4}\sim \frac{8}{\pi \hbar}
\label{eq:refined-top}
\end{eqnarray}
\begin{figure}[htb]
\center
\includegraphics[scale=0.5]{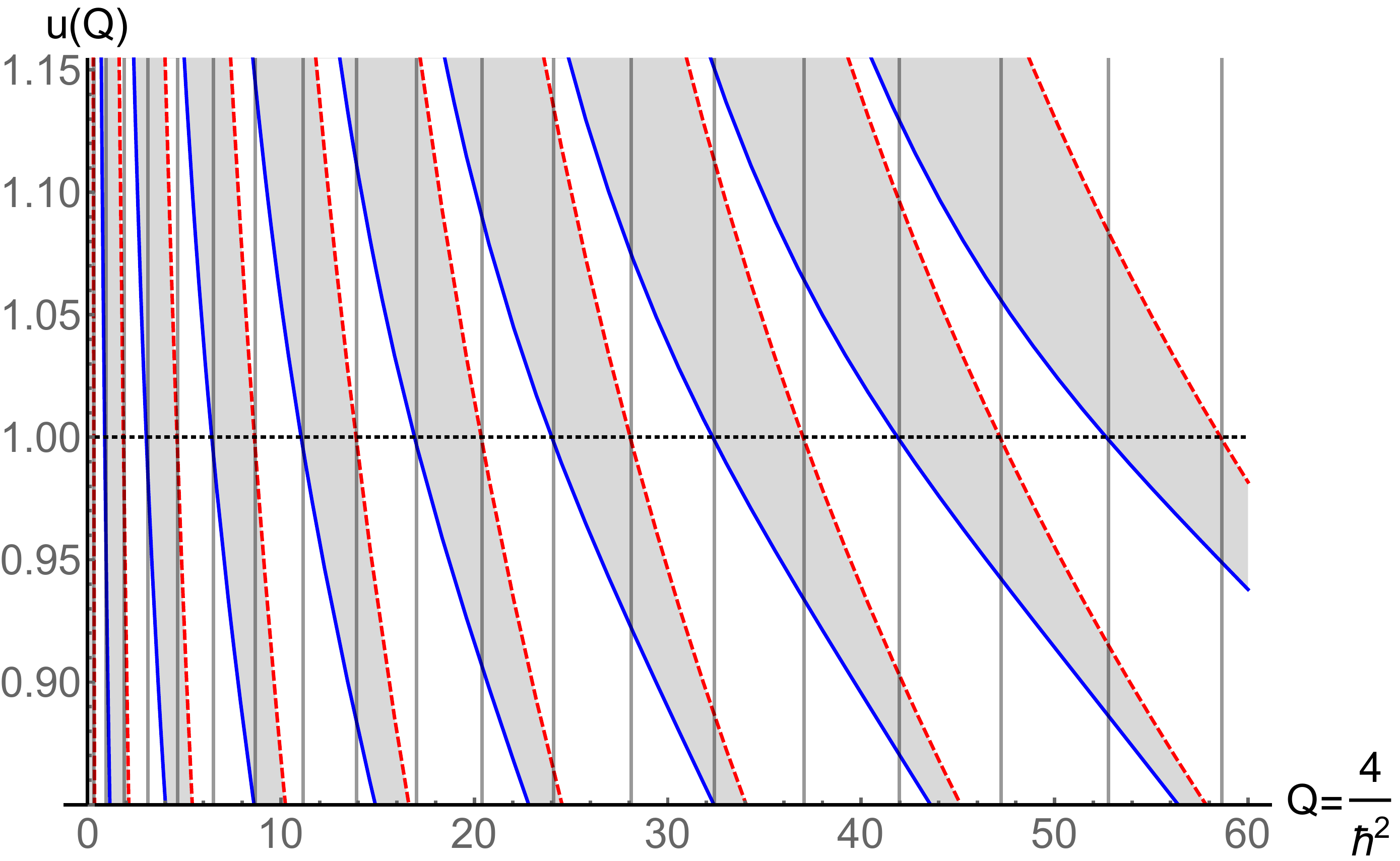}
\caption{Plots of the bands (shaded) and gaps (unshaded) in the magnetic region, in the vicinity of the barrier top where $u=1$. The vertical lines mark the values $Q=\frac{\pi^2}{16}\left(N\pm \frac{1}{4}\right)^2$, where $N$ is the band label, and which coincide very accurately with the points at which the band/gap edges intersect the line $u=1$.}
\label{fig:f6}
\end{figure}
as shown in Figure \ref{fig:f6}. We also see that in the immediate vicinity of these points, the dependence of 
$u$ on $Q (\equiv \frac{4}{\hbar^2})$ is approximately linear, with a slope depending inversely quadratically on $N$:
\begin{eqnarray}
u^{(\pm)}_N \sim 1-\frac{c_N}{\left(N\pm \frac{1}{4}\right)^2}\left(Q-\frac{\pi^2}{16}\left(N\pm \frac{1}{4}\right)^2\right)\quad, \quad c_N\sim O(1)
\label{eq:linear}
\end{eqnarray}
This implies that in the vicinity of the barrier top, where $u\approx 1$, the bands and gaps are of equal width, and are not exponentially narrow. This can be seen clearly in Figure \ref{fig:f6}, where 
from a given intersection point at $u=1$, we observe equal widths of the band and gap above and below that intersection point. In fact, 
\begin{eqnarray}
\Delta u^{\rm band}\sim \Delta u^{\rm gap}\sim O(\hbar)
\label{eq:leading}
\end{eqnarray}

Another way to understand the smooth transition across the barrier top is to note that in both the dyonic and electric region, we must avoid poles by analytically continuing $\hbar$ (or equivalently $g$) off the positive real axis, which effectively gives a small imaginary part to $u$. This then avoids the divergences of $a_n(u)$ in the region $u\sim 1$, in a way that connects smoothly. Since all the actions are expressed in terms of the elliptic functions ${\mathbb E}$ and ${\mathbb K}$, this relies on the analytic continuation properties of these functions. 
More precisely, the analytical continuation properties connect different points in the moduli space as follows
\bea
\begin{pmatrix} a_0(-u)\\  a^D_0(-u) \end{pmatrix} =-i \begin{pmatrix} \pm1 & 0 \\  1 & \pm1 \end{pmatrix} \begin{pmatrix} a_0(u) \\  a^D_0(u) \end{pmatrix} 
  \ea
where $\pm1$ denotes $sign(\mc Im[u])$. This is the manifestation of the residual $\mathbb Z_2$ symmetry of the broken $U(1)_R$ symmetry of the gauge theory \cite{Bilal:1996sk}. For the subleading terms from \eqref{eq:ans-general} we see that 
\bea
\begin{pmatrix} a_n(-u)\\  a^D_n(-u) \end{pmatrix} =-(-1)^n\, i \begin{pmatrix} \pm1& 0 \\  1 & \pm1 \end{pmatrix} \begin{pmatrix} a_n(u) \\  a^D_n(u) \end{pmatrix}\,.
\label{eq:z2}
\ea 
In particular, for $-1\leq u\leq 1$, this relation can be used to relate the magnetic regime ($u\sim1$) to the dyonic regime ($u\sim-1$). The appearance of $ sign(\mc Im[u])$ reflects the necessity of analytic continuation mentioned above, and is ultimately related with the fact that $\hbar \in \mathbb R^+$ is a Stokes line for the small $\hbar$ expansion. 

Also, the region with $0\leq u\leq  1$,  can be related to the $1\leq u\leq \infty$ region by invoking the inversion relations for the elliptic functions ${\mathbb E}$ and ${\mathbb K}$. These elliptic functions have non-trivial connection formulas, with ambiguous non-perturbative imaginary contributions when evaluated on the real line. See, for example, \url{http://dlmf.nist.gov/19.7.E3} 
\begin{eqnarray}
{\mathbb K}\!\left(\frac{1}{k^2}\right)&=&k\, \left({\mathbb K} \mp i\, {\mathbb K}^\prime \right)
\\
{\mathbb E}\!\left(\frac{1}{k^2}\right)&=&\frac{1}{k} \left( {\mathbb E}\pm i\, {\mathbb E}^\prime-{k^{{
\prime}}}^{2}{\mathbb K}\mp ik^{2}\, {\mathbb K}^\prime\right)
\end{eqnarray}
where $\pm$ is determined by the sign of $Im(k^2)$.\footnote{Note that these analytic continuation properties are incorrectly stated, without the $\pm$ signs, in many books and tables \cite{fettis}.}
With the above connection formulas, the transition across the barrier is smooth.

\subsection{Connecting Strong and Weak Coupling Regimes}
\label{sec:strong-weak}

In this section we explain the connection between the large $\hbar$ and small $\hbar$ expansions. These two expansions are very different in nature: the former is convergent and the latter is divergent. Yet the exact WKB methods described in the previous section can be used to relate one to the other. We present an explicit method for generating the small $\hbar$ expansion for the low lying modes (i.e. $\lambda \ll 1$), which governs the dyonic region, using the knowledge acquired from the ``continued fraction expansion'' for $\lambda \gg 1$, which governs the electric region.

In the electric region, the energy eigenvalue $u$ for high lying states can be obtained from the continued fraction expression 
\bea
2 Q u-N^2 -{Q^2\over 2 Q u -(N-2)^2-}{Q^2\over 2 Q u -(N-4)^2-}\dots={Q^2 \over (N+2)^2- 2 Q u -}{Q^2 \over (N+4)^2- 2 Q u -}\dots \nonumber\\
\label{eq:cont-frac}
\ea
as an expansion in $Q^2 = 16/\hbar^4$, by making a series ansatz for $u$ and equating both sides of \eqref{eq:cont-frac}. This expansion is valid for high lying levels with $N \gg 1/\hbar$. The first few terms of it are given in \eqref{eq:smallq}. Next, we substitute $a$ for $N$ by using the relation $a=N \hbar /2$, which is the exact quantization condition \eqref{eq:electric-quant} in the electric region. In the electric region $a =\lambda/2 \gg 1$ and we reorganize $u$ as a series in $1/a^4$ and $\hbar$, as written in \eqref{eq:smallq-u}. Having obtained this double series, we invert it and obtain $a$ as a double series in $u$ and $\hbar$, i.e. $a(u, \hbar)=\sqrt{2u}\sum_{n,k}c_{n,k}\hbar^{2n}u^{-2k}$, which is valid for $u\gg1$. So far, this is an alternative way of obtaining the large $\lambda$ expansion \eqref{eq:electric-wkb-expansion}, directly from the continued fraction expression (\ref{eq:cont-frac}), complementary to the WKB derivation  decribed earlier in this section. 

However, from the general form of the WKB period $a(u,\hbar)$ given in \eqref{eq:ans-general}, we know that each coefficient of $\hbar^{2n}$ is related to the lowest order coefficient [with $n=0$] via a differential operator, with expansion coefficients $\kappa^{(n)}_k$. Therefore, for any given order $\hbar^{2n}$, by applying the differential operator in \eqref{eq:ans-general} to the coefficient of $\hbar^0$ of \eqref{eq:electric-wkb-expansion}, and comparing with the coefficient of $\hbar^{2n}$, we can solve for $\kappa^{(n)}_k$s. Once we know these coefficients, $\kappa^{(n)}_k$, we can construct both $a^D (u)$ and $a(u)$ from $a^D_0(u)$ and $a_0(u)$, \textit{for any u}! 

To obtain the small $\hbar$, small $N$ (dyonic) expansion for $u$, we then expand $a$ and $a^D$ around $u\sim-1$. In the dyonic region, the exact quantization condition is ${\mc R}e[a]=a+a^D=\hbar/2(N+1/2)$. Then we invert back to obtain $u({\mc R}e[a],\hbar)$ which is the perturbative, small $\hbar$ expansion. The non-perturbative, instanton induced part of the trans-series can be obtained from $a^D$ as described in Section \ref{sec:wkb-dyonic}.

Before closing this section, we also would like to point out that the continued fraction expansion of $u(a,\hbar)$ given in \eqref{eq:smallq} can alternatively be obtained from the conformal block expansion obtained from the AGT correspondence \cite{Alday:2009aq,Fateev:2009aw} in the $\epsilon_2\rightarrow0$ limit.  The instanton part of the Nekrasov partition function\footnote{Note that our overall normalization of ${\cal F}$ differs from some of the literature on the subject.} has a rather simple group theoretical expansion given as \cite{Gaiotto:2009ma,Mironov:2009uv}
\bea
Z^{inst.}_{Nek.}(a;\epsilon_1,\epsilon_2)=\exp\left(-{4 \pi i\over \epsilon_1\epsilon_2} {\cal F}(a;\epsilon_1,\epsilon_2)\right)=\sum_{n=0}^\infty\left({\Lambda^2 \over \epsilon_1 \epsilon_2}\right)^{2n} Q^{-1}_\Delta([1^n],[1^n])
\ea
 where $Q_\Delta(Y,Y^\prime)=\langle\Delta|L_{Y}\,L_{-Y^\prime}|\Delta\rangle$ is the Shapovalov matrix associated with a conformal primary $|\Delta\rangle$ with Young tableaux $Y$ and $Y^\prime$ satisfying $|Y|=|Y^\prime|$, and $[1^n]$ is a shorthand notation for $Y=\{1,1,\dots,1\}$ with $|Y|=n$. The first couple of relevant entries of the inverse of $Q_\Delta$ are 
 \bea
Q^{-1}_\Delta([1],[1])={1\over 2 \Delta}\quad,\quad Q^{-1}_\Delta([11],[11])={8\Delta +c \over 4 \Delta (16\Delta^2 +2 c\Delta-10\Delta+c) }\quad,\quad\dots
 \ea 
The higher order terms can be computed in a straightforward way. The AGT correspondence maps the conformal dimension, $\Delta$, and central charge, $c$, to the gauge theory variables: 
\bea
\Delta={1\over\epsilon_1\epsilon_2}\left(a^2-{(\epsilon_1+\epsilon_2)^2\over 4}\right)\quad,\quad c=1-{6(\epsilon_1+\epsilon_2)^2\over\epsilon_1\epsilon_2}\,.
\ea
In the $\epsilon_2\rightarrow0$ limit, the prepotential stays finite:
\bea
{\cal F}^{inst.}_{NS}(a;\epsilon_1)=-{\epsilon_1\over 4 \pi i }\lim_{\epsilon_2\rightarrow0}\epsilon_2 \log\left(Z^{inst.}_{Nek.}(a,\epsilon_1,\epsilon_2)\right)
\ea
Using Matone's relation \eqref{eq:Matone} and switching from $a$ to $N$ using the exact quantization we get
\bea
{i\pi\over2}\Lambda {\partial \mc F^{inst.}_{NS}\over \partial \Lambda} &=&-{\epsilon_1\over 16}\lim_{\epsilon_2\rightarrow0}\,\epsilon_2\, {\partial \over \partial\log\Lambda }\log\left(\sum_{n=0}^\infty\left({\Lambda^2 \over \epsilon_1 \epsilon_2}\right)^{2n} Q^{-1}_\Delta([1^n],[1^n])\right)\nonumber\\
&=&{\hbar^2\over8}\left(\frac{8 \Lambda ^4}{\left(N^2-1\right) \hbar ^4}+\frac{8 \Lambda ^8 \left(5 N^2+7\right)}{\left(N^2-4\right) \left(N^2-1\right)^3 \hbar ^8}+\dots\right)
\ea
which exactly coincides with the expansion \eqref{eq:smallq} we obtained from the continued fraction \eqref{eq:cont-frac}. The leading term, $N^2$, comes from the perturbative part of the Nekrasov expansion.

\section{Worldline Instantons and Multi-photon Vacuum Pair Production}
\label{sec:wli}

As discussed in Section \ref{sec:wkb}, despite the fact that the perturbative expansions in the electric region are convergent, there are still non-perturbative instanton effects, leading to narrow gap splittings  high in the spectrum. These splittings (\ref{eq:electric-gap-width}) have a very different form from those  (\ref{eq:dyonic-band-width}) low in the spectrum. 
This is analogous to the results of Drukker, Mari\~no and Putrov \cite{Drukker:2010nc} for the large $N$ expansion of the ABJM matrix model, in which non-perturbative effects are related to complex space-time instantons.
Here we give an analogous physical identification with {\it worldline instantons} \cite{Dunne:2005sx}, for this effect in the Mathieu spectrum. Furthermore, we present a simple physical interpretation of the transition between different spectral regions, in terms of the transition from tunneling pair production to multi-photon pair production. This  analogy gives a physical example in which a non-perturbative quantity, the pair production probability, turns smoothly into a multi-photon expression of a very different form.

 Recall the semiclassical result of Br\'ezin and Itzykson \cite{brezin}, and Popov \cite{popov}, (generalizing Keldysh's work in atomic ionization \cite{keldysh}), that for a monochromatic time dependent electric field, ${\mathcal E}(t)={\mathcal E}\,\cos(\omega\, t)$, the usual Schwinger formula \cite{Heisenberg:1935qt,Schwinger:1951nm,Dunne:2004nc} for the vacuum pair production probability in a background electric field becomes
\begin{eqnarray}
\text{probability} \sim
\exp\left[-\frac{m^2\, \pi}{{\mathcal E}}\, g(\gamma)\right]
\label{eq:prob}
\end{eqnarray}
Here the dimensionless ``Keldysh adiabaticity parameter'' is defined as
\begin{eqnarray}
\gamma\equiv \frac{m\, \omega}{{\mathcal E}}
\label{eq:adiabaticity}
\end{eqnarray}
For this monochromatic time-dependent  field ${\mathcal E}(t)={\mathcal E}\,\cos(\omega\, t)$, one finds for the function $g(\gamma)$ in (\ref{eq:prob}) the expression \cite{brezin,popov,Dunne:2004nc,Dunne:2005sx}:
\begin{eqnarray}
g(\gamma)&=&\frac{4}{\pi}\, \frac{\sqrt{1+\gamma^2}}{\gamma^2}\left[ {\mathbb K}\left(\frac{\gamma^2}{1+\gamma^2}\right)- {\mathbb E}\left(\frac{\gamma^2}{1+\gamma^2}\right)\right]
\label{eq:gfactor} \\
&\sim&
\begin{cases}
1-\frac{1}{8}\gamma^2 +\dots \quad, \quad \gamma\ll 1 \cr
\frac{4}{\pi\, \gamma}\, \ln (4 \gamma) +\dots \quad, \quad \gamma\gg 1
\end{cases}
\label{eq:glimits}
\end{eqnarray}
In the static limit, $\gamma\to 0$, we recover the familiar Heisenberg-Schwinger result, 
\begin{eqnarray}
\text{probability} \sim
\exp\left[-\frac{m^2\, \pi}{{\mathcal E}}\right]
\label{eq:schw}
\end{eqnarray}
which is obviously non-perturbative in the strength of the applied field, and has a well-known interpretation in terms of tunneling from the Dirac sea through the barrier created by the constant electric field. In the opposite limit of a high frequency field, where $\gamma\gg 1$, the logarithmic beahvior of $g(\gamma)$, shown in (\ref{eq:glimits}), implies that the semiclassical  result (\ref{eq:prob})  in fact produces a perturbative result:
\begin{eqnarray}
\text{probability}\sim
\exp\left[-\frac{m^2\, \pi}{{\mathcal E}}\, g(\gamma)\right] \to \left(\frac{{\mathcal E}}{4 m\, \omega}\right)^{4m/\omega}
\label{eq:multiphoton}
\end{eqnarray}
The physical interpretation of (\ref{eq:multiphoton})  is that $2m/\omega$ is the multi-photon number, the number of photons of energy $\omega$ required to excite the virtual pair over the ``binding energy'' $2m c^2$, by a multi-photon process, rather than by tunneling  \cite{brezin,popov,keldysh}. It's a probability, so the power is twice this multi-photon number. The final answer is perturbative, as it is a power of the applied electric field, but it is a very high power in the relevant semiclassical limit where $\omega\ll m$. Thus we see that the semiclassical expression (\ref{eq:prob}) has two very different limits, and we show below that this is analogous to the transition from  exponentially narrow bands  (\ref{eq:dyonic-band-width}) low in the Mathieu spectrum, to power-law narrow gaps (\ref{eq:electric-gap-width}) high  in the Mathieu spectrum.

First, recall that the above pair production results have a natural interpretation in terms of {\it worldline instantons} \cite{Dunne:2005sx}, which are finite action periodic solutions to the classical Euclidean equations of motion for the world line of a charged particle in the background electric field:
\begin{eqnarray}
\ddot{x}_\mu= F_{\mu\nu}(x)\dot{x}_\nu
\label{eq:wli}
\end{eqnarray}
For such worldline instanton solutions, a semiclassical approximation to the Feynman's world line path integral representation of the QED effective action gives the leading expression for the pair production probability
\begin{eqnarray}
\text{probability} \sim
\exp\left[-S[x_{\rm instanton}]\right]\qquad, \qquad S[x]=\int \left(\frac{1}{2} \dot{x}^2+A_\mu \dot{x}_\mu\right)d\tau
\label{eq:prob2}
\end{eqnarray}
For the cosine electric field, ${\mathcal E}(t)={\mathcal E}\,\cos(\omega\, t)$, the  classical trajectories are known, and the associated worldline instanton action indeed equals the exponent in (\ref{eq:prob}, \ref{eq:gfactor}) \cite{Dunne:2005sx}. 

The relation to the Mathieu equation is the following. The pair production probability can alternatively be computed via a Bogoliubov transformation as a quantum mechanical reflection coefficient in the Klein-Gordon equation (we take zero electron/positron momentum, to get the leading effect, and treat scalar QED instead of spinor QED, since this also gives the same exponential behavior) \cite{brezin,popov}:
\begin{eqnarray}
-\ddot{\phi}-\left(m^2+\left(\frac{{\mathcal E}}{\omega}\, \sin(\omega \, t)\right)^2\right)\phi=0\quad\to\quad
\phi^{\prime\prime}+\left[\left(\frac{m}{\omega}\right)^2+\frac{1}{2}\left(\frac{{\mathcal E}}{\omega^2}\right)^2\right]\phi-\frac{1}{2}\left(\frac{{\mathcal E}}{\omega^2}\right)^2\, \cos(2x)\phi=0
\end{eqnarray}
written now in Mathieu form. So we identify the Mathieu equation parameters:
\begin{eqnarray}
A=\left[\left(\frac{m}{\omega}\right)^2+\frac{1}{2}\left(\frac{{\mathcal E}}{\omega^2}\right)^2\right]\qquad , \qquad Q= \left(\frac{{\mathcal E}}{2\omega^2}\right)^2 
\end{eqnarray}
Therefore, we further identify [using the scalings in (\ref{hbar}) and (\ref{hbar-g})]
\begin{eqnarray}
\hbar \equiv \frac{2}{\sqrt{Q}}=\frac{4\omega^2}{{\mathcal E}}  \qquad, \qquad  u\equiv \frac{A}{2Q}=1+2\gamma^2
\label{eq:hbar-schwinger}
\end{eqnarray}
Further, note that the leading dual action $a_0^D$ in (\ref{aperiods}) can be re-expressed as 
\begin{eqnarray}
a_0^D(u)&=&
 -{2i\over \pi} \left({\mathbb E}\left(\frac{1-u}{2}\right)-\frac{1}{2} (u+1)\, {\mathbb K}\left(\frac{1-u}{2}\right)\right) \\
&=& 2 i\sqrt{2} \sqrt{u+1} \left({\mathbb K}\left(\frac{u-1}{u+1}\right)-{\mathbb E}\left(\frac{u-1}{u+1}\right)\right)
\label{aperiods2}
\end{eqnarray}
Therefore,
\begin{eqnarray}
2\left(\frac{2 \pi}{\hbar} Im\left[a_0^D(u)\right] \right)\quad \longleftrightarrow \quad \frac{m^2 \pi}{{\mathcal E}}\, g(\gamma)
\label{eq:ident1}
\end{eqnarray}
This implies that\footnote{Note that the pair production probability is related to a bounce, which is a closed-path instanton/anti-instanton configuration, referred to as a ``worldline instanton''; while the Mathieu band or gap splitting is a single-instanton effect. For  time-dependent electric fields associated with more realistic laser pulses, there are important quantum interference effects and these are captured by complex instanton trajectories \cite{Dumlu:2010ua}.}
\begin{eqnarray}
(\text{gap width})^2 \sim \exp\left[-2\left(\frac{2 \pi}{\hbar} Im\left[a_0^D(u)\right] \right)\right] \quad \longleftrightarrow \quad  \text{probability} \sim \exp\left[-\frac{m^2 \pi}{{\mathcal E}}\, g(\gamma)\right]
\label{eq:ident2}
\end{eqnarray}
To see how this works in the various limits, recall that  in the semiclassical approach to vacuum pair production, the electron mass sets the dominant scale, so we require ${\mathcal E}\ll m^2$ and $\omega\ll m$. But this still permits arbitrary values of the adiabaticity parameter $\gamma\equiv m\omega/{\mathcal E}$. In the static limit, $\gamma\ll 1$, this implies $u\sim 1$, and 
\begin{eqnarray}
\frac{4\pi}{\hbar} Im\left[a_0^D(u)\right] \sim \frac{2\pi}{\hbar} (u-1) \sim \frac{4\pi \gamma^2}{\hbar} =\frac{\pi m^2}{{\mathcal E}} 
\label{eq:ident3}
\end{eqnarray}
which leads to the standard Schwinger formula for pair production in a static electric field.

On the other hand, in the multi-photon limit, $\gamma\gg 1$, we have 
\begin{eqnarray}
\frac{m}{\omega} \gg \frac{{\mathcal E}}{\omega^2} 
\end{eqnarray}
so we can consistently consider the hierarchy of scales
\begin{eqnarray}
{\mathcal E}\ll \omega^2\ll m\, \omega \ll m^2
\end{eqnarray}
Then defining  $N\equiv \frac{m}{\omega}$, we see that it corresponds to the gap label index $N$ in the Mathieu spectral problem. Indeed, in this limit
\begin{eqnarray}
N\equiv \frac{m}{\omega} \gg \frac{{\mathcal E}}{\omega^2}\equiv \frac{4}{\hbar} 
\end{eqnarray}
in Mathieu language. Thus, from the Mathieu equation gap splitting formula (\ref{eq:smallq-splitting}) for large $N$ [using Stirling's formula], the identification (\ref{eq:hbar-schwinger}) leads to
\begin{eqnarray}
(\text{gap width})^2\sim \left(\frac{e}{N\,\hbar}\right)^{4N} \sim \left(\frac{e\, {\mathcal E}}{4 m\, \omega}\right)^{4m/\omega} 
\label{analogy}
\end{eqnarray}
which, up to a multiplicative pre-factor, is the Br\'ezin/Itzykson/Popov multi-photon expression (\ref{eq:multiphoton}). 

\section{A Simple Proof of the Dunne-\"Unsal Relation and its Geometric Interpretation}

In this Section we present a proof of the Dunne-\"Unsal relation \eqref{eq:magic} from the gauge theory point of view. In terms of quantum mechanics, this relation remarkably links the perturbative fluctuations around the vacuum to the non-perturbative fluctuations around instantons \cite{Dunne:2013ada}, as mentioned in Section \ref{sec:electric}.  In addition to the SUSY inspired proof, we also show that the Dunne-\"Unsal relation can be identified as the generalization of the Picard-Fuchs equation for the quantized elliptic curve with nonzero $\hbar$. Physically this identification is an explicit example of the connection between the resurgent trans-series expansion and the geometry of compact Riemann surfaces.

Our starting point of the proof is the generalization of Matone's relation for $\hbar\neq0$,
\bea
u(a,\hbar)={i\pi\over2}\Lambda {\partial \mc F_{NS}(a,\hbar)\over \partial \Lambda} -{\hbar^2 \over 48}\,. 
\label{eq:matone2}
\ea
where as mentioned in Section \ref{sec:electric}, the instanton part of the original relation \cite{Matone:1995rx} is unchanged and the shift in $u$ is due to the perturbative contribution to $\mc F_{NS}(a,\hbar)$ \cite{Gorsky:2014lia}. The next step is to observe that the the prepotential can be expressed as follows
\bea
\mc F_{NS}(a,\hbar):=\Lambda^2  \mc {\hat F_{NS}}\left({a\over \Lambda},{\hbar \over \Lambda} \right):=\Lambda^2 \mc {\hat F_{NS}}\left(\hat a,\hat \hbar \right)
\ea
where the we use the symbol  $\hat\,$  to denote dimensionless quantities.\footnote{Note that $\hbar$ has mass dimension $1$ in the gauge theory/QM correspondence. This can easily be seen from the Schr\"odinger equation \eqref{eq:schrodinger}.} This follows from rescaling the 
Schr\"odinger equation \eqref{eq:schrodinger} and the corresponding periods \eqref{eq:BSperiods}.  After the rescaling, it follows that
\bea
\Lambda {\partial \mc F_{NS}(a,\hbar)\over \partial \Lambda}&=&  2\Lambda ^2 \mc {\hat F_{NS}}-\Lambda^2 \hat a {\partial \mc {\hat F_{NS}}(\hat a,\hat\hbar)\over \partial \hat a}-\Lambda^2 \hat \hbar {\partial \mc {\hat F_{NS}}(\hat a,\hat\hbar)\over \partial \hat \hbar}\\
&=& 2\mc F_{NS}(a,\hbar)-a {\partial \mc F_{NS}(a,\hbar)\over \partial a}-\hbar{\partial \mc F_{NS}(a,\hbar)\over \partial \hbar}\,.
\label{eq:scaling}
\ea
Then, by differentiating \eqref{eq:matone2} with respect to $a$ and using \eqref{eq:scaling} we obtain
\bea
{2 i\over\pi}{\partial u (a,\hbar)\over \partial a}+a^D(a,\hbar)-a \,{\partial a^D(a,\hbar)\over \partial a}-\hbar{\partial a^D(a,\hbar)\over \partial\hbar}=0\,,
\label{eq:magic-periods}
\ea 
where we used the fact that $\partial \mc F/\partial a=a^D$.  The final step in order to arrive at the form which relates the instanton expansion  to perturbative expansion is to switch to the variables $A_{ZJJ}(B,\hbar)$ (which encodes the fluctuations around instantons) and $E_{ZJJ}(B,\hbar)$ (which encodes the perturbation expansion) where $E_{ZJJ}=(u-1)/\hbar$, $B=2 a/\hbar$, and $A_{ZJJ}$ is defined as in \eqref{a}. In these variables, \eqref{eq:magic-periods} becomes  
\begin{eqnarray}
\frac{\partial E_{\rm ZJJ}}{\partial B}=- {\hbar\over 16} \left(2B+\hbar\frac{\partial A_{\rm ZJJ}}{\partial \hbar}\right)
\end{eqnarray}

It is also illuminating to switch the independent variables from $(a,\hbar)$ to $(u,\hbar)$. With this change of variables, the terms in \eqref{eq:magic-periods} transform as follows:
\bea
{\partial u \over \partial a } =\left({\partial a \over \partial u }\right)^{-1}\quad,\quad {\partial a^D \over \partial a} =\left({\partial a \over \partial u }\right)^{-1} {\partial a^D \over \partial u} \quad,\quad
{\partial a^D\ \over \partial \hbar} =
\left({\partial a \over \partial u }\right)^{-1}  \left({\partial a^D \over \partial \hbar}{\partial a\over \partial u}-{\partial a \over \partial \hbar}{\partial a^D\over \partial u}\right)\,
\ea
 where the independent variables on the left-hand side are $(a,\hbar)$ and on the right-hand side are $(u,\hbar)$. Then, the Dunne -\" Unsal relation takes the form of Picard-Fuchs equation, extended to nonzero $\hbar$:
\bea
\left(a-\hbar{\partial a\over \partial\hbar}\right){\partial a^D\over \partial u}-\left(a^D-\hbar{\partial a^D\over \partial \hbar}\right){\partial a\over \partial u}={2 i\over\pi}\,.
\label{eq:pf}
\ea 
Note that \eqref{eq:pf} is invariant under the $SL(2,\mathbb Z)$ transformations that act on the pair $(a,a^D)$ as expected. It is further useful to express \eqref{eq:pf} in terms of the expansions \eqref{eq:periods}:
\bea
a_0 {d a^D_0 \over d u}-a^D_0 {d a_0 \over d u}&=&{2i\over \pi}\\
\label{eq:pf-exp1}
\sum_{k=0}^n(1-2k)\left(a_k {da^D_{n-k}\over du}-a^D_k {da_{n-k}\over du}\right)&=&0,\quad n\geq1\,.
\label{eq:pf-exp2}
\ea
The second line of this equation  shows that all the higher order terms contribute as zero to the Picard-Fuchs equation, and the constant $2i/\pi$ on the right hand side does not get any corrections at nonzero $\hbar$. In other words, by using the Riemann bilinear identities on \eqref{eq:pf-exp1} and \eqref{eq:pf-exp2} we deduce that only the zeroth order periods $a_0$, $a^D_0$ contribute to the ``total flux'' on the torus, whereas the higher order terms do not. However, the cancellation of the flux from higher order terms is a result of rather intricate cancellations that involve contributions from different orders as seen in \eqref{eq:pf-exp2}.

The identification of the Dunne-\"Unsal relation with the Picard-Fuchs equation provides an explicit geometric interpretation of the connection between perturbative series and instanton expansion. The perturbative corrections to the energy eigenvalue are characterized by the quantization of the real period, while the exponentially suppressed instanton corrections are characterized by the dual pure imaginary period. The Picard-Fuchs relation connects these two independent periods. Furthermore, even though the perturbative and non-perturbative expansions take very different forms in different regions of the energy spectrum, this connection holds throughout the spectrum. 
 
It is interesting to note that similar connection formulas between perturbative and non-perturbative physics exist for other QM potentials, such as the double-well and SUSY double-well \cite{Dunne:2013ada}. In fact, in general a relation like (\ref{eq:magic})  exists whenever the spectrum is such that the associated Riemann surface is described by just the two dual actions, $a$ and $a^D$. For more general cases, where the Riemann surface is of higher genus, the relation will generalize to include multiple pairs of actions whose number is equal to the genus \cite{bdu}. 

\section{Lam\'e Equation and  $SU(2)$ ${\mathcal N}=2^*$ SUSY Gauge Theory}
\label{sec:lame}

As discussed in Section \ref{sec:wkb} for the Mathieu equation, a formal expansion of the action may be obtained by an all-orders WKB analysis. This can then be expanded in the high or low energy region. Instead of repeating these steps for the Lam\'e system, we summarize the novel features that arise in the dyonic region in Section \ref{sec:lame-dyonic}, and then use a completely different technique
to study the electric  region, using the relation between WKB and the KdV hierarchy: see Section \ref{sec:lame-electric}.

\subsection{Resurgent Analysis in the Dyonic Region}
\label{sec:lame-dyonic}

The dyonic gauge region of the $\mathcal N=2^*$ theory corresponds to the $\lambda\ll 1$ spectral region of the Lam\'e system. Previous analyses \cite{He:2010xa} have not taken into account the fact that in this region the spectral expansions are divergent and non-Borel-summable, as for the Mathieu system, and so should be described by a resurgent trans-series. In fact, the associated resurgent structure is even richer than for the Mathieu system, due to the existence of both real and complex instantons \cite{Basar:2013eka}. This is ultimately due to the fact that the Lam\'e potential is doubly-periodic in the complex plane, and even though the quantum mechanical path integral is a sum over {\it real} path configurations, the existence of {\it complex}  saddle paths has a direct influence on the divergent structure of perturbation theory \cite{Basar:2013eka}.

The Jacobian form of the Lam\'e equation is conventionally written \url{http://dlmf.nist.gov/29.2.i}:
\begin{eqnarray}
\frac{{d}^{2}\psi}{{dz}^{2}}+(H-\nu(\nu+1)k^{2}{\mathop{\mathrm{sn}\/}\nolimits^{%
2}}\left(z,k^2\right))\psi=0
\label{eq:lame}
\end{eqnarray}
The Jacobi elliptic function ${\rm sn}^2$ is related to the  Weierstrass ${\mathcal P}$-function as\footnote{We use the standard period conventions of \cite{nist}; other papers sometimes use differing conventions \cite{He:2010xa,KashaniPoor:2012wb,Billo:2013fi,Billo:2014bja}.}:
\begin{eqnarray}
{\mathcal P}\left(x+i\, {\mathbb K}^\prime; {\mathbb K},i\, {\mathbb K}^\prime\right)=-\frac{1}{3}\left(k^2+1\right)+k^2 \, {\rm sn}^2(x; k^2)
\label{eq:psn}
\end{eqnarray}
where ${\mathbb K}^\prime\equiv {\mathbb K}(1-k^2)$. 
The Lam\'e equation (\ref{eq:lame})  reduces to the Mathieu system in a special scaling limit: since ${\rm sn}^2(z; 0)=\sin^2(z)$, we must combine the $k^2\to 0$ limit with the $\nu\to \infty$ limit, in such a way that the combination $\kappa^2\equiv \nu(\nu+1)\, k^2$ remains finite. This then leads to the natural identifications:
\begin{eqnarray}
\hbar\leftrightarrow \frac{4}{\sqrt{\nu(\nu+1)k^2}}\quad, \quad u\leftrightarrow -1+\frac{\hbar^2}{8}\, H
\label{eq:lame-id}
\end{eqnarray}
With these identifications, we can express the standard perturbative expansions of the Lam\'e eigenvalues (\url{http://dlmf.nist.gov/29.7.E1}) as:
\begin{eqnarray}
u(N, \hbar) &=&-1+\hbar\left(N+\frac{1}{2}\right) -\frac{\hbar^2}{16}(k^2+1)\left[\left(N+\frac{1}{2}\right)^2+\frac{1}{4}\right] \\
&-&\frac{\hbar^3}{2^8}\left[(1+k^2)^2 \left(\left(N+\frac{1}{2}\right)^3+\frac{3}{4}\left(N+\frac{1}{2}\right)\right)-4k^2 \left(\left(N+\frac{1}{2}\right)^3+\frac{5}{4}\left(N+\frac{1}{2}\right)\right)\right]-\dots 
\nonumber
\end{eqnarray}
Notice that we recover the Mathieu expression (\ref{eq:largeq}) in the limit $k^2\to 0$.

For a given $N\ll 1/\hbar$ and $k^2$, these expansions are divergent, with factorially growing coefficients. But the large order behavior can be alternating or non-alternating, depending on $k^2$.
This more intricate structure is due to the existence of both real and complex instantons, and is discussed in detail in \cite{Basar:2013eka}.\footnote{The paper \cite{Basar:2013eka} uses another form of the elliptic potential, ${\rm sd}^2(z, k^2)$, which has a manifest symmetry under $k^2\to 1-k^2$. This potential is related to the ${\rm sn}^2(z, k^2)$ by a simple Landen transformation.} In particular, it means that the leading large-order growth of the perturbative coefficients is not solely governed by the real instanton/anti-instanton action, but also by the complex (`ghost') instanton/anti-instanton action.

Associated with this resurgent divergent structure of the perturbative expansions is the existence of non-perturbative band splittings, for any $k^2$. Approximate expressions for these band splittings are given in (\url{http://dlmf.nist.gov/29.7.E5}); for details see \cite{muller-lame,volkmer}. Translating these results to our notation (\ref{eq:lame-id}),  the splitting of the $N^{\rm th}$ band is
\begin{eqnarray}
\Delta u_N^{\rm band}\sim&& \\
&&\hskip -1.5cm \frac{2\hbar}{N!}  \sqrt{\frac{2}{\pi}}
\left(\frac{32}{\hbar (1-k^2)}\right)^{N+\frac{1}{2}}
\left(\left(\frac{1-k}{1+k}\right)^{\frac{1}{k}}\right)^{\frac{4}{\hbar}}
\left[1-\frac{\hbar}{32}(1+k^2)\left(3\left(N+\frac{1}{2}\right)^2+4\left(N+\frac{1}{2}\right)+\frac{3}{4}\right)-\dots \right]
\nonumber
\end{eqnarray}
This band-splitting is a one-instanton effect, as the instanton action for the Lam\'e potential is \cite{Dunne:1999zc,Dunne:2002at}
\begin{eqnarray}
\frac{1}{\hbar} S_{\rm inst}=\frac{4}{\hbar\, k}\ln\left(\frac{1+k}{1-k}\right)
\label{eq:lameinst}
\end{eqnarray}
Note that this reduces to the Mathieu expression (\ref{eq:bandwidth}) as $k^2\to 0$, recalling that $(1-k)^{1/k}\to 1/e$ in this limit. 

\subsection{WKB, Gelfand-Dikii Expansion, and KdV, in the Electric Region}
\label{sec:lame-electric}

As discussed in Section \ref{sec:wkb} for the Mathieu equation, the electric region of the Lam\'e spectrum can be studied by using a large $u$ expansion of the all-orders WKB actions $a(u, \hbar)$ and $a^D(u, \hbar)$. 
Here we describe a different, complementary, technique to analyze the electric region, using the
work of Gelfand and Dikii concerning the high-energy asymptotic expansion of the resolvent \cite{gelfand-dikii,perelomov}. The coefficients of this expansion are expressed directly in terms of the KdV integrals, evaluated on the QM potential. This observation is particularly interesting for the Lam\'e equation because the expansion coefficients are simple polynomials, with interesting combinatorial properties, in the scaling parameter that multiplies the elliptic potential \cite{grosset}. This is described below.

For the Schr\"odinger equation with a  periodic potential $V(x)$ of period $L$,
\begin{eqnarray}
-\frac{d^2}{dx^2} \psi+V(x)\, \psi=E\, \psi
\label{eq:kdv-schrodinger}
\end{eqnarray}
the Gelfand-Dikii expansion \cite{gelfand-dikii,perelomov} relates the level number $N$ to the high energy asymptotics as
\begin{eqnarray}
\frac{\pi N}{L} \sim  \sqrt{E}-\frac{1}{L} \sum_{j=0}^\infty \frac{I_{j+1}[V]}{(4 E)^{j+1/2}}\qquad, \quad E\to + \infty
\label{eq:gd}
\end{eqnarray}
where $I_{j+1}[V]$ are functionals of the potential $V(x)$ given by the KdV conserved quantities. These can be generated by simple recursion relations, and the first few are:
\begin{eqnarray}
I_1[V]&=&\int_0^L V\, dx\\
I_2[V]&=&\int_0^L V^2\, dx\\
I_3[V]&=&\int_0^L \left((V^\prime)^2+2\, V^3\right)\, dx \\
I_4[V]&=&\int_0^L \left(\left(V^{\prime\prime}\right)^2+10\, V\, \left(V^\prime\right)^2+5\, V^4\right)\, dx \quad, \quad \dots
\end{eqnarray}
It is an instructive exercise to compute these KdV integrals for a constant potential, and also for the Mathieu potential, to confirm the all-orders WKB analysis in Section \ref{sec:wkb}. For example, for a constant potential, $V=V_0$, 
\begin{eqnarray}
\frac{N \pi}{L}\sim\sqrt{E}-\frac{V_0}{\sqrt{4E}}-\frac{V_0^2}{(4E)^{3/2}}-\frac{2V_0^3}{(4E)^{5/2}}-\frac{5V_0^4}{(4E)^{7/2}}-\dots
\label{eq:kdv-check1}
\end{eqnarray}
which is the $E\to +\infty$ expansion of the exact result $\sqrt{E-V_0}$.
For the Mathieu system, to compare the Mathieu equation (\ref{mathieu1}) with the Gelfand-Dikii form (\ref{eq:kdv-schrodinger}), we identify $V \leftrightarrow \frac{2}{\hbar^2} \cos x$, $E\leftrightarrow \frac{2}{\hbar^2} \, u$, and  $L\leftrightarrow 2\pi$. The KdV integrals are simple to evaluate:
\begin{eqnarray}
&&I_1\left[\frac{2}{\hbar^2} \cos x\right] =0\quad, \quad I_2\left[\frac{2}{\hbar^2} \cos x\right] =\pi \left(\frac{2}{\hbar^2}\right)^2\quad, \quad I_3\left[\frac{2}{\hbar^2} \cos x\right] =\pi \left(\frac{2}{\hbar^2}\right)^2 \nonumber\\
&&  I_4\left[\frac{2}{\hbar^2} \cos x\right] =\pi \left(\frac{2}{\hbar^2}\right)^4\left(\frac{15}{4}+\left(\frac{\hbar^2}{2}\right)^2\right)\quad, \quad \dots
\label{eq:trig-faulhaber}
\end{eqnarray}
which leads to the expansion:
\begin{eqnarray}
\frac{N \hbar}{2}&\sim& \sqrt{2u}-\frac{1}{4}\frac{1}{(2u)^{3/2}} -\frac{1}{16} \frac{1}{(2u)^{5/2}}\, \hbar^2-\frac{1}{64} \frac{1}{(2u)^{7/2}}\left(\hbar^4+15\right)-\dots
\label{eq:wkb-electric-2}
\end{eqnarray}
Note that the coefficients of the inverse  powers of $u^{1/2}$ are {\it polynomials} in $\hbar^2$, of increasing order. This expansion (\ref{eq:wkb-electric-2}) is a re-arrangement of the all-orders WKB expansion in (\ref{eq:electric-wkb-expansion}). So, inverting (\ref{eq:wkb-electric-2}) to express $u$ as a function of $\hbar$ and $a\equiv N \hbar/2$, we recover the expansion (\ref{eq:smallq-u}) for $u$ in the electric regime where $a\gg 1$.

Turning now to the Lam\'e system, Grosset and Veselov \cite{grosset} considered the potential $V= 2\mu {\mathcal P}$, where $\mathcal P$  is the Weierstrass ${\mathcal P}$-function (\ref{eq:psn}), and $\mu$ is a multiplicative scaling parameter to be specified below.  Grosset and Veselov showed that the KdV integrals $I_{j+1}[V]$ reduce to simple polynomials in $\mu$, with coefficients expressed in terms of the Weierstrass invariants $g_2$ and $g_3$, and $\eta_1/\mathbb K$, where $\eta_1\equiv \zeta(\mathbb K)$ \cite{grosset}:
\begin{eqnarray}
I_{j+1}[2\,\mu\, \mc P]\equiv F_{j+1}(\mu)
\label{eq:faulhaber}
\end{eqnarray}
These polynomials, $F_{j+1}(\mu)$,\footnote{Not to be confused with the prepotential which we denote with the calligraphic letter $\cal F$.} have interesting combinatorial properties, and were named the {\it elliptic Faulhaber polynomials} \cite{grosset}\footnote{The name derives from a  remarkable connection between the classical Faulhaber polynomials of Number Theory and the KdV integrals for soliton-like potentials \cite{fairlie}, generalized to elliptic functions associated with periodic arrays of solitons \cite{grosset}. In this sense, the polynomials in (\ref{eq:trig-faulhaber}) could be referred to as {\it trigonometric Faulhaber polynomials}.}. 
Thus, for this Weierstrassian form of the Lam\'e equation, (\ref{eq:kdv-schrodinger}) with $V(x)={\mathcal P}\left(x+i\, {\mathbb K}^\prime; {\mathbb K},i\, {\mathbb K}^\prime\right)$, the high-energy Gelfand-Dikii expansion (\ref{eq:gd}) can be written
\begin{eqnarray}
\frac{\pi N}{2 {\mathbb K}} \sim  \sqrt{E}- \frac{1}{2 \mathbb K}\sum_{j=0}^\infty \frac{F_{j+1}(\mu)}{(4 E)^{j+1/2}}\qquad, \quad E\to +\infty
\label{eq:gd2}
\end{eqnarray}
In order to facilitate the comparison with results in the physics literature concerning the Nekrasov prepotential in ${\mathcal N}=2^*$ theories \cite{KashaniPoor:2012wb,Billo:2013fi,Billo:2014bja}, we rewrite the coefficients of the elliptic Fauhaber polynomials in terms of the Eisenstein series
\bea
E_2(\tau)=-{2 \pi i\over \zeta(2)}{\partial\over \partial\tau} \log(\eta(\tau))\quad,\quad E_k(\tau)={1\over 2\zeta(k)} \sum_{(n,m)\in \mathbb Z^2\setminus(0,0)}{1\over(n+\tau\,m)^k},\quad (k>2).
\ea
where  $\eta(\tau)$ is the Dedekind eta function, and $E_2$, $E_4$ and $E_6$ can also be written as:
\begin{eqnarray}
E_2=3\left(\frac{2\mathbb K}{\pi}\right)^2 \, \frac{\eta_1}{\mathbb K}
\quad, \quad
E_4=\frac{3}{4} \left(\frac{2\mathbb K}{\pi}\right)^2 \, g_2
\quad, \quad 
E_6=\frac{27}{8} \left(\frac{2\mathbb K}{\pi}\right)^2 \, g_3
\end{eqnarray}
Thus, all coefficients of the elliptic Faulhaber polynomials are expressed in terms of just the first few Eisenstein series $E_2$, $E_4$ and $E_6$.
In our conventions for the periods of the Weierstrass function, the modular parameter $\tau$ is identified as
\bea
\tau\equiv{i \mathbb K^\prime\over \mathbb K}\,.
\ea
For notational simplicity, we  suppress the $\tau$ (therefore $k^2$) dependence of the Eisenstein series in the following equations.
 
Then  the first few elliptic Faulhaber polynomials $F_{j+1}(\mu)$ are
\begin{eqnarray}
{1\over 2 \mathbb K} F_{1}(\mu)&=& -\left({\pi \over 2 \mathbb K}\right)^2 \frac{1}{3} [E_2]\, (2\mu)\\
{1\over 2 \mathbb K} F_{2}(\mu)&=& \left({\pi \over 2 \mathbb K}\right)^4\frac1{9} [E_4]\, (2\mu)^2\\
{1\over 2 \mathbb K} F_{3}(\mu)&=& \left({\pi \over 2 \mathbb K}\right)^6\frac{2}{135} \left(\left[4E_6-9 E_2 E_4\right] (2\mu)^3 -12\left[E_6-E_2 E_4\right] (2\mu)^2\right)\\
{1\over 2 \mathbb K} F_{4}(\mu)&=& \left({\pi \over 2 \mathbb K}\right)^8\frac{1}{189} \left( 
\frac{5}{3} \left[15 E_4^2-8 E_2 E_6\right] (2\mu)^4
-80  \left[E_4^2-E_2 E_6\right] (2\mu)^3
+96 \left[E_4^2-E_2 E_6\right] (2\mu)^2
\right)\nonumber \\
\end{eqnarray}
Further properties of these elliptic Faulhaber polynomials $F_{j+1}(\mu)$ are discussed in \cite{grosset}.

The Gelfand-Dikii expansion (\ref{eq:gd2}) can be viewed as a high-energy expansion of an all-orders Bohr-Sommerfeld relation, which can be inverted\footnote{It is interesting to note that the Lagrange inversion of series can be naturally expressed in terms of Young tableaux \cite{stanley,Langmann:2014rja}.} to yield
\begin{eqnarray}
E\sim \left({\pi \over 2\,\mathbb K }\right)^2\left(N^2 +\sum_{j=0}^\infty \frac{{\cal G}_{j+1}(\mu)}{N^{2j}}\right)
\label{eq:gd3}
\end{eqnarray}
where the polynomials ${\cal G}_{j+1}(\mu)$ are simple combinations of the elliptic Faulhaber polynomials $F_{j+1}(\mu)$:
\begin{eqnarray}
{\cal G}_1(\mu)&=& -\frac{1}{3} [E_2]\, (2\mu) \\
{\cal G}_2(\mu)&=& \frac{1}{36}  \left[E_4-E_2^2\right] (2\mu)^2 \\
{\cal G}_3(\mu)&=& \frac{1}{540} \left( 
\left[2E_6+3E_2 E_4-5 E_2^3\right] (2\mu)^3
-6 \left[E_6-E_2 E_4\right] (2\mu)^2 \right)\\
{\cal G}_4(\mu)&=& \frac{1}{9072} \left( \left[-35 E_2^4+7 E_2^2 E_4+10E_4^2+18E_2 E_6\right] (2\mu)^4
+12 \left[-5 E_4^2-2 E_2 E_6+7 E_2^2 E_4\right] (2\mu)^3 \right. \nonumber\\
&&\left.
\hskip1cm+72\left[-E_2 E_6+E_4^2\right] (2\mu)^2 \right)
\end{eqnarray}
It is convenient to define the rescaled and shifted eigenvalue, absorbing the $j=0$ term from the sum:
\bea
\tilde u\equiv  \frac{1}{2} \left(\frac{\hbar}{2}\right)^2 \left(\left(\frac{2\mathbb K}{\pi}\right)^2 E
+\frac{E_2}{3} (2\mu)\right)
\label{eq:utilde}
\ea 
Then in terms of the action variable $a=\hbar N/2$,   (\ref{eq:gd3}) becomes
\begin{eqnarray}
\tilde u\sim \frac{1}{2} a^2+\frac{1}{2} \sum_{j=1}^\infty \left(\frac{\hbar}{2}\right)^{2j+2}
\frac{{\mathcal G}_{j+1}(\mu)}{a^{2j}}
\label{eq:gd4}
\end{eqnarray}
It is  a non-trivial check that with the scaling of (\ref{eq:lame-id}) and (\ref{eq:psn}), we identify $2\mu\leftrightarrow \frac{16}{\hbar^2\, k^2}$, and in the $k^2\to 0$ limit we find that (\ref{eq:gd4}) does indeed reduce to the Mathieu large period expansion in (\ref{eq:smallq-u}).

We can now compare the Bohr-Sommerfeld energy expansion (\ref{eq:gd4}) with the expansion of the Nekrasov-Shatashvili prepotential for the $SU(2)$ ${\mathcal N}=2^*$ theory 
\cite{Dorey:1996ez,Minahan:1997if,KashaniPoor:2012wb,Billo:2013fi,Billo:2014bja}\footnote{Note that our normalization of ${\cal F}$ differs from the normalization in \cite{KashaniPoor:2012wb} as ${\cal F}_{\rm here}=-{1\over 4 \pi i}{\cal F}_{\rm there}$.}:
\begin{eqnarray}
{\cal F}_{NS}^{{\cal N}=2^*}(a, \hbar, m)\sim {1\over 2} \tau \, a^2-{{\cal H}_0\over 2\pi i} \log(\eta(\tau))-{1\over 4\pi i} \sum_{j=1}^\infty {{\cal H}_{j}\over 2^{j+1}\,j\,a^{2j}}
\label{eq:nsstar}
\end{eqnarray}
where ${\cal H}_0$ is expressed in terms of the scalar mass, $m$, and the Nekrasov deformation parameter in the Nekrasov-Shatashvili limit $\hbar=\epsilon_1$ as
\begin{eqnarray}
{\cal H}_0\equiv m^2-\frac{\hbar^2}{4}
\label{eq:hzero}
\end{eqnarray}
The higher coefficients ${\cal H}_{j}$  are polynomials in ${\mathcal H}_0$:
\begin{eqnarray}
{\cal H}_1({\cal H}_0)&=& \frac{1}{12} [E_2] {\cal H}_0^2 \\
{\cal H}_2({\cal H}_0) &=& \frac{1}{360}\left(\left[5 E_2^2+E_4\right] {\cal H}_0^3 -3\,  [E_4]\, {\cal H}_0^2\, \hbar^2 \right)\\
{\cal H}_3({\cal H}_0) &=& \frac{1}{60480}\left(\left[175 E_2^3+84 E_2 E_4+11 E_6\right] {\cal H}_0^4
-36 \left[7 E_2 E_4+3 E_6\right] {\cal H}_0^3 \hbar^2
+180 [E_6]\, {\cal H}_0^2 \hbar^4 \right)
\nonumber\\
\label{eq:hs}
\end{eqnarray}
Using the differentiation properties of the Eisenstein series 
\begin{eqnarray}
\hskip-0.4cm{1\over 2\pi i}\frac{d}{d\tau} E_2=\frac{1}{12}\left[E_2^2-E_4\right]\quad, \quad 
{1\over 2\pi i}\frac{d}{d\tau}E_4=\frac{1}{3}\left[E_2 E_4-E_6\right]\quad, \quad
{1\over 2\pi i}\frac{d}{d\tau}E_6={1\over 2}\left[E_2 E_6-E_4^2\right]
\label{eq:dedq}
\end{eqnarray}
and the identification 
\begin{eqnarray}
2\, \mu \leftrightarrow \frac{{\cal H}_0}{\hbar^2}  \equiv  \frac{m^2}{\hbar^2}-\frac{1}{4}
\label{eq:id}
\end{eqnarray}
we observe the remarkable fact that the expansion coefficients ${\mathcal H}_{j}$ are directly related to the polynomials ${\mathcal G}_{j+1}$:
\begin{eqnarray}
\frac{1}{2\pi i} \frac{\partial}{\partial \tau} {\mathcal H}_j =-\frac{j}{2^{j+1}} \hbar^{2j+2} \, {\mathcal G}_{j+1}\left(\mu\right) \quad, \quad {\rm with}\quad 2\mu\leftrightarrow \frac{{\mathcal H}_0}{\hbar^2}
\end{eqnarray}
We thus arrive at the Matone relation for the ${\mathcal N}=2^*$ theory:
\begin{eqnarray}
\tilde u =\frac{\partial}{\partial\tau} {\mathcal F}_{NS}^{{\cal N}=2^*}
+\frac{E_2}{24}\left(m^2-\frac{\hbar^2}{4}\right)
\label{eq:matone3}
\end{eqnarray}
The shift in $\tilde{u}$ is just the shift in (\ref{eq:utilde}), required to have a smooth reduction to the Mathieu eigenvalue, as the $\mathcal N=2^*$ theory reduces to the $\mathcal N=2$ theory in the $m^2\to\infty$ limit, which is combined with $k^2\to 0$ with $m^2 k^2$ finite.\footnote{Effectively, we are subtracting the average value of the Weierstrass potential over one period, so that the first KdV integral, $I_1[V]$, vanishes, as it does for the Mathieu potential.} 

It is interesting to note that when the scalar mass $m$ is related to the Nekrasov deformation parameter 
$\epsilon_1\equiv \hbar$ as
\begin{eqnarray}
m=\left(j+\frac{1}{2}\right)\hbar \quad \Rightarrow\quad 2\mu=j(j+1)
\label{eq:fg}
\end{eqnarray}
then if $j$ is an integer the Lam\'e system becomes {\it finite gap} \cite{dubrovin}, with the number of gaps equal to $j$. In this case, much more explicit descriptions of the spectral information can be given \cite{dubrovin,grosset}, and there is also a precise strong/weak band/gap duality symmetry \cite{Dunne:2002at}. Notably for integer values of $j$, the associated density of states $\rho(E) dE$ has an algebraic geometric meaning: it is an abelian differential of the second kind over a genus-$j$ Riemann surface. Therefore the quantization condition $\pi N\sim\int \rho(E)dE$ carries a clear geometric interpretation.

Thus, in the large action spectral region of the Lam\'e equation we identify the all-orders WKB Bohr-Sommerfeld expansion with the Gelfand-Dikii expansion, and the eigenvalue is identified with the scalar condensate moduli parameter of the  $\mathcal N=2^*$  gauge theory. But, as in the Mathieu equation, this Bohr-Sommerfeld expression (\ref{eq:gd4}) for the energy is only part of the story. Identifying the action $a$ with $N\hbar/2$, the Bohr-Sommerfeld expression gives a (convergent) perturbative expression for the location of the $N^{th}$ gap high in the spectrum. However, physically it is clear that there are also non-perturbatively small gaps in the spectrum, and these are related to instantons. As in the Mathieu case discussed in Section 3, these gaps arise due to poles in the expansion coefficients: see the continued fraction expressions at \url{http://dlmf.nist.gov/29.3.iii}. Since the structure is quite similar, we do not repeat all the steps here. Further discussion of the Lam\'e system is deferred to a future publication.

\section{Conclusions}

In this paper we have applied resurgence and all-orders exact WKB to provide a complete description of the different spectral regions of the Mathieu and Lam\'e systems, stressing their close connection with the low energy behavior of ${\mathcal N=2}$ SUSY SU(2) gauge theories. Defining a 't Hooft parameter, $\lambda=N\, \hbar$, where $N$ is the spectral level number, we associate the large $\lambda$ regime with the gauge electric regime, the small $\lambda$ regime with the gauge dyonic regime, and the  $\lambda\sim 1$ regime with the gauge magnetic regime. We have shown that exact WKB provides a complete description of all regimes, including all non-perturbative effects, and permits direct mappings between these sectors.  Previous analyses based on all-orders Bohr-Sommerfeld relations described only part of the information, neglecting non-perturbative band or gap splittings. This analysis also shows that the relation between perturbative and non-perturbative contributions is radically different in the different regimes. The familiar relation between divergent perturbative expansions and non-perturbative effects must be generalized to accommodate the large $\lambda$ regime where perturbative expansions are convergent, and non-perturbative effects are associated with poles of expansion coefficients. This generalizes the resurgent trans-series analysis of \cite{zjj,Dunne:2013ada} to large $\lambda$. We provide a physical analogy of this change of non-perturbative behavior, in the transition from tunneling pair production to multi-photon pair production in the Schwinger effect in a time-dependent electric field, as the adiabaticity of the field changes. This physics is naturally described by worldline instantons. Since the Nekrasov deformation parameters have the physical interpretation of constant graviphoton fields, the SUSY gauge theory significance of these worldline instantons, and their associated non-perturbative effects, should be further understood.

The reinterpretation of the spectral problem in the language of the Nekrasov partition function leads naturally to a simple proof of a quantum mechanical resurgence relation \cite{Dunne:2013ada} that shows that all non-perturbative information of the trans-series eigenvalues is subtly encoded within perturbation theory, which is an extreme form of resurgence behavior. We also give a geometrical interpretation of this fact in terms of the Riemann bilinear identity and the Picard-Fuchs equation, complementary to work on quantum geometry \cite{Mironov:2009uv,KashaniPoor:2012wb,Krefl:2013bsa} and Whitham dynamics \cite{Gorsky:2014lia}. We demonstrate a direct relation between the all-orders Bohr-Sommerfeld relation and the Gelfand-Dikii expansion, based on the KdV invariants, and show explicitly how the ${\mathcal N}=2^*$ theory reduces to the ${\mathcal N}=2$ system. Future work will develop these ideas further \cite{bdu}. 
\vskip 1cm

{\bf Acknowledgements. }
We thank C. Bender, O. Costin, A. Gorsky, M. Mari\~no, N. Nekrasov, M. \"Unsal and A. Vainshtein for discussions and correspondence. GB thanks the DOE for support through grant DE-FG02-93ER-40762. GD acknowledges support from DOE grant DE-FG02-13ER41989, and acknowledges hospitality and support from the Physics Department at the Technion, and the Theoretical Physics Institute at Friedrich-Schiller Universit\"at, Jena, where much of this work was done. GD also thanks the DFG for a Mercator Professor Grant, and the US-German Fulbright Commission for a US Senior Scholar Award.

\appendix
\section{Simple Analog of Uniform Asymptotic Behavior}

As discussed in \cite{Stone:1977au,Dunne:2012ae}, certain features of the divergence of the perturbative expansion for the ground state energy of the Mathieu system are captured by the zero-dimensional partition function
\begin{eqnarray}
Z(g)\equiv \frac{1}{2\pi}\int_{-\pi}^\pi dx\, e^{-\frac{1}{g}\, \cos x} &=& I_0\left(\frac{1}{g}\right)\nonumber\\
&\sim & \frac{1}{\sqrt{2\pi/g}}\, e^{\frac{1}{g}}\sum_{n=0}^\infty \frac{\Gamma\left(n+\frac{1}{2}\right)^2}{\pi\, 2^n n!} g^n \quad, \quad g\to 0^+
\label{eq:zero}
\end{eqnarray}
The relation to the quantum mechanical spectral problem arises because this zero-dimensional partition function gives the resummed leading derivative expansion contribution to the heat kernel expansion ${\rm tr}\, e^{-H t}$, from which the resolvent and spectrum can be extracted. The perturbative expansion in (\ref{eq:zero}) is asymptotic, with non-alternating factorial large order behavior of the expansion coefficients, $c_n\sim (n-1)!/\pi$. The series is non-Borel-summable and should instead be represented by a trans-series, which has just two exponential terms since $Z(g)$ satisfies a second order differential equation \cite{Costin:2009}:\footnote{For the Lam\'e potential, $Z(g)$ satisfies a third-order differential equation, and this fact is reflected in the appearance of three different exponential terms, which can be identified with the perturbative vacuum and both real and complex ('ghost') instantons \cite{Basar:2013eka}.}
\begin{eqnarray}
Z(g)
&\sim & \frac{1}{\sqrt{2\pi/g}}\, e^{\frac{1}{g}}\sum_{n=0}^\infty \frac{\Gamma\left(n+\frac{1}{2}\right)^2}{\pi\, 2^n n!} g^n \pm  \frac{i}{\sqrt{2\pi/g}}\, e^{-\frac{1}{g}}\sum_{n=0}^\infty (-1)^n \frac{\Gamma\left(n+\frac{1}{2}\right)^2}{\pi\, 2^n n!} g^n
\label{eq:zero-trans}
\end{eqnarray}
where $-\frac{\pi}{2}+\delta \leq \mp {\rm ph}\, g\leq \frac{3}{2}\pi-\delta$.
This trans-series expression contains all information about the function, including the Stokes phenomenon. In particular, the second sub-dominant part is required in order to be consistent with  the analytic continuation connection formula for the Bessel functions:
\begin{eqnarray}
K_0(e^{\pm i\pi} z) =K_0(z)\mp i\,\pi\, I_0(z)
\label{eq:bessel-connection}
\end{eqnarray}
Note that in the large $g$ limit this function has a convergent ``strong-coupling'' expansion:
\begin{eqnarray}
Z(g)\sim \sum_{k=0}^\infty \frac{1}{\left(2^k\, k!\right)^2}\, \frac{1}{g^{2k}}
\quad, \quad g\to +\infty
\label{eq:strong}
\end{eqnarray}
The dependence on the level number $N$ can be modeled by considering instead
\begin{eqnarray}
Z_N(g)\equiv  I_N\left(\frac{1}{g}\right)
\sim  \frac{1}{\sqrt{2\pi/g}}\, e^{\frac{1}{g}}\sum_{n=0}^\infty c_n(N) \, g^n \quad, \quad g\to 0^+\quad (N\,\,{\rm fixed})
\label{eq:Nzero}
\end{eqnarray}
where the series is again divergent and non-Borel-summable, with $c_n(N)\sim (-1)^N \frac{(n-1)!}{\pi}$, at large perturbative order $n$, as in (\ref{eq:zero}). But we can also consider the large $N$ limit with $g$ fixed:
\begin{eqnarray}
Z_N(g)
\sim  
\frac{1}{\sqrt{2\pi N}}\left(\frac{e}{2\, N\, g}\right)^N \quad, \quad N\to +\infty \quad (g\,\,{\rm fixed})
\label{eq:Nzerog}
\end{eqnarray}
which has the same form as the gap splitting (\ref{eq:smallq-splitting}) high in the spectrum where $N\gg 1/g$. On the other hand, there is also a {\it uniform} expansion where $N\to\infty$ and $g\to 0$ such that $N g$ is kept fixed. The uniform expansion valid for all values of the 't Hooft parameter $\lambda\equiv N\, g$ is:
\begin{eqnarray}
Z_N\left(\frac{1}{g}\right)&=&I_N\left(N\, \frac{1}{N g}\right)\nonumber\\
&\sim&  \frac{1}{\sqrt{2\pi}}\, \frac{\exp\left[\sqrt{N^2+\frac{1}{g^2}}\right]}{\left(N^2+\frac{1}{g^2}\right)^{1/4}}
\left(\frac{\frac{1}{N\, g}}{1+\sqrt{1+\frac{1}{(N\, g)^2}}}\right)^N
 \sum_{n=0}^\infty \frac{1}{N^n}\, U_n\left(\frac{1}{\sqrt{1+\frac{1}{(N\, g)^2}}}\right)  \nonumber\\
 && \quad\quad ,\qquad N\to +\infty \quad (0<N\, g<\infty)
 \label{eq:Nzero-uniform}
\end{eqnarray}
where $U_n$ is a (known) polynomial of degree $3n$ [\url{http://dlmf.nist.gov/10.41.E3}]. This uniform expression interpolates smoothly between the two extreme limits and describes the function in the intermediate region, the analog of the ``magnetic region''.

\section{WKB coefficients}
\label{app:kappas}

In this appendix we list the first five set of coefficients $\kappa^{(n)}_k$, that enter into the relation
 \bea
  a_n(u)=\sum_{k=0}^n \kappa^{(n)}_k u^k {d^{n+k} a_0(u)\over du^{n+k}}\,.
  \ea
   that connects the higher order WKB cycles to the leading order one. $n$ denotes the coefficient of $\hbar^{2n}$ in the WKB expansion. The method for calculating them is explained in Section \ref{sec:strong-weak}.
  \begin{center}
\begin{tabular}{ c|cccccc}
  \hline
 $n\downarrow/ k\rightarrow$   &0 &1 &2 &3  &4&5  \\ \hline
     $1$          &1/48 &1/24 &\,&\,  &\, &\, \\ \hline
    $2$          &5/1536 &1/192 &7/5760 &\,  &\,   &\,  \\ \hline
     $3$          &41/57344 &153/143360 &79/215040 &31/967680  &\, &\,  \\ \hline
    $4$           &${15229\over70778880}$ &${9539\over30965760}$ &${517\over4128768}$ &${13\over716800}$  &${127\over154828800}$ &\,  \\ \hline
     $5$           &${484249\over5813305344}$&${5049503\over43599790080}$ &${8430053\over163499212800}$ &${780341\over81749606400}$  &${61729\over81749606400}$  &${73\over3503554560}$ \\ 
   \hline
\end{tabular}
\end{center}

\end{document}